\documentclass[aps,prd,preprint,groupedaddress,showpacs]{revtex4-1}

\usepackage{amsmath,amssymb}
\usepackage[dvips]{graphicx}
\usepackage{braket}

\begin{document}

%\preprint{}

\preprint{OU-HET 683/2010}

\title{Entanglement Entropy of Two Black Holes and\\ Entanglement Entropic Force}


\author{Noburo Shiba}
\email{shiba@het.phys.sci.osaka-u.ac.jp }
%{Your e-mail address}
%\thanks{}
\affiliation{Department of Physics, Graduate School of Science,
Osaka University, Toyonaka, Osaka 560-0043, Japan}


\date{\today}

\begin{abstract}
We study the entanglement entropy, $S_C$, of the massless free scalar field on the outside region $C$ of two black holes $A$ and $B$ whose radii are $R_1$ and $R_2$
and how it depends on the distance, $r(\gg R_1,R_2)$ , between two black holes. 
If we can consider the entanglement entropy as thermodynamic entropy, 
we can see the entropic force %(we call this entanglement entropic force)
 acting on the two black holes from the $r$ dependence of $S_C$.
%We consider the case that the state of the massless free scalar field is the vacuum state which depend how to choose the time coordinate.
%We choose the coordinate system which covers whole space time and does not have the coordinate singularity on the horizons.
We develop the computational method based on that of Bombelli et al to obtain the $r$ dependence of $S_C$ of scalar fields whose Lagrangian is quadratic with respect to the scalar 
fields.
%p
First we study $S_C$ in $d+1$ dimensional Minkowski spacetime.
In this case the state of the massless free scalar field is the Minkowski vacuum state
and we replace two black holes by two imaginary spheres, and we take the %partial 
trace over the degrees of freedom residing in the imaginary spheres.
We obtain the leading term  of $S_C$ with respect to $1/r$.
The result is $S_C=S_A+S_B+\tfrac{1}{r^{2d-2}} G(R_1,R_2)$, where  $S_A$ and $S_B$ are the entanglement entropy on the inside region of $A$ and $B$, 
and $G(R_1,R_2) \leq 0$.
We do not calculate  $G(R_1,R_2)$ in detail, but we show how to calculate it. 
%p
In the black hole case we use the method used in the Minkowski spacetime case with some modifications.
We show that $S_C$ can be expected to be the same form as that in the Minkowski spacetime case.
But in the black hole case, $S_A$ and $S_B$ depend on $r$, so we do not fully obtain the $r$ dependence of $S_C$. 
%p
Finally we assume that the entanglement entropy can be regarded as thermodynamic entropy, %we can consider the entanglement entropy as thermodynamic entropy
and %as a result 
consider the entropic force acting on two black holes.
We argue how to separate  the entanglement entropic force from other force 
and how to cancel $S_A$ and $S_B$ whose $r$ dependence are not obtained.
Then we obtain the physical prediction which can be tested experimentally in principle.

%$r$ dependence 
%First we study the entanglement entropy, $S_C$, of the massless free scalar field on the outside region $C$ of two imaginary spheres $A$ and $B$ whose radii are $R_1$ and $R_2$
%and how it depends on the distance, $r$ , between two spheres. 

%We obtain 

%We study the entanglement entropy, $S_C$, of two black holes whose radii are $R_1$ and $R_2$ 
%and how it depends on the distance, $r$ , between two black holes.  
%We develop the computational method based on that of Bombelli et al to obtain the $r$ dependence of $S_C$ of scalar fields whose Lagrangian is quadratic with respect to the scalar fields.
%We assume that we can consider the entanglement entropy as thermodynamic entropy
%and as a result the entropic force acts on two black holes.
%We consider three situations to separate  the entanglement entropic force from other force,
%and obtain the physical prediction which can be tested experimentally in principle.
% and discuss the possibility to measure the entanglement entropic force.

%the distance ($r$) between  dependence of and 
%We 
%We develop the method to calculate  
\end{abstract}
%03.65.Ud, Entanglement and quantum nonlocality (e.g. EPR paradox, Bell's inequalities, GHZ states, etc.) (for entanglement production and manipulation, see 03.67.Bg; for entanglement measures, witnesses etc., see 03.67.Mn; for entanglement in Bose-Einstein condensates, see 03.75.Gg)
%04.70.Dy,Quantum aspects of black holes, evaporation, thermodynamics
% 11.90.+t Other topics in general theory of fields and particles (restricted to new topics in section 11)


%\pacs{11.90.+t}  %check   11.25.Tq 
%\keywords{}

\maketitle

\section{\label{sec level1}Introduction}

%\subsection{}
%\subsubsection{}

Entanglement entropy in quantum field theory (QFT) was originally studied to explain the black hole entropy \cite{Bombelli:1986rw,Srednicki:1993im}.
Entanglement entropy is generally defined as the von Neumann entropy $S_A=-Tr \rho_A \ln \rho_A$ corresponding to the reduced density matrix $\rho_A$ of a subsystem $A$.
When we consider quantum field theory in $d+1$ dimensional spacetime $\mathbb{R}\times N$, where $\mathbb{R}$ and $N$ 
denote the time direction and the $d$ dimensional space-like manifold respectively,
%In quantum field theory in $d+1$ dimensional spacetime 
we define the subsystem by a $d$ dimensional domain 
$A\subset N$ at fixed time $t=t_0$.
(So this is also called geometric entropy.)
Entanglement entropy naturally arises when we consider the black hole because we cannot % access the inside region of it.
obtain the information in the black hole.
In fact,  in the vacuum state the leading term of the entanglement entropy of $A$ is proportional to the area of the boundary $\partial A$ in many cases 
\cite{Bombelli:1986rw,Srednicki:1993im}.
This is similar to the black hole entropy, and extensive 
studies have been carried out \cite{Hawking:2000da, Kabat:1995eq, Susskind:1994sm, Frolov:1993ym, Jacobson:1994iw, 'tHooft:1984re}.
%\cite{Hawking:2000da,Kabat:1995eq,Susskind:1994sm,Frolov:1993ym,Jacobson:1994iw,Frolov:1993ym,'tHooft:1984re}.

In this paper we study the entanglement entropy, $S_C$, of the massless free scalar field on the outside region $C$ of two black holes $A$ and $B$ whose radii are $R_1$ and $R_2$
and how it depends on the distance, $r(\gg R_1,R_2)$ , between two black holes. 
We consider the case that the state of the massless free scalar field is the vacuum state which depends how to choose the time coordinate.
We choose the coordinate system which covers whole space time and does not have the coordinate singularity on the horizons.
If we can consider the entanglement entropy as thermodynamic entropy, 
we can see the entropic force (we call this entanglement entropic force) acting on the two black holes from the $r$ dependence of $S_C$.

%In this paper we study the entanglement entropy, $S_C$ , of the outside region of two black holes (A and B) whose radii are $R_1$ and $R_2$ 
%and how it depends on the distance, $r$ , between two black holes. 
%If we can consider the entanglement entropy as thermodynamic entropy, 
%we can see the entropic force (we call this entanglement entropic force) acting on the two black holes from the $r$ dependence of $S_C$.

In Section \ref{general} we obtain the general behavior of the entanglement entropy of two disjoint regions in translational invariant vacuum in general QFT.
In Section \ref{review} we review the computational method of entanglement entropy in free scalar fields \cite{Bombelli:1986rw}. 
There are some computational methods of entanglement entropy \cite{Calabrese:2004eu,Holzhey:1994we,Ryu:2006bv} .
See \cite{Ryu:2006ef,Casini:2009sr} for reviews.
That of Bombelli et al \cite{Bombelli:1986rw} is most straightforward and powerful enough to obtain the $r$ dependence of $S_C$ in free scalar field theory.
In Section \ref{flat} 
we study $S_C$ in $d+1$ dimensional Minkowski spacetime.
In this case the state of the massless free scalar field is the Minkowski vacuum state
and we replace two black holes by two imaginary spheres, and we take the %partial 
trace over the degrees of freedom residing in the imaginary spheres.
We develop the method of Bombelli et al and obtain the leading term of $S_C$ with respect 
to $1/r$. % for a free massless scalar field in $d+1$ dimensional Minkowski spacetime.
The result in this section agrees with the general behavior in Section \ref{general}.
This method can be used for any scalar fields in curved space time whose Lagrangian is quadratic 
with respect to the scalar fields (i.e higher derivative terms can exist).
In Section \ref{BH} we consider the black hole case. We use the method used in Section \ref{flat} with some modifications.
We show that $S_C$ can be expected to be the same form as that in the Minkowski spacetime case.
But in the black hole case $S_A$ and $S_B$ depend on $r$, so we do not fully obtain the $r$ dependence of $S_C$. 
In Section \ref{prediction} 
we assume that the entanglement entropy can be regarded as thermodynamic entropy, %we can consider the entanglement entropy as thermodynamic entropy,
and consider the entanglement entropic force.
We argue how to separate  the entanglement entropic force from other force 
and how to cancel $S_A$ and $S_B$ whose $r$ dependence are not obtained.
Then we obtain the physical prediction which can be tested experimentally in principle,
and discuss the possibility to measure the entanglement entropic force.
In Appendix B we obtain a formula for a finite series as a by-product of our calculation.

%In general $S_{AB} =S_C$ when the total system $ABC$ is pure state.
%We consider the vacuum state, so  $S_{AB} =S_C$ .

\section{general behavior} \label{general}

We consider entanglement entropy of two disjoint regions ($A$ and $B$) in translational invariant vacuum in general QFT 
in $d+1$ dimensional spacetime $(d\geq 2)$.
We will show that $S_C$ reaches its maximum value when $r\rightarrow \infty $.
%First, we 

There are several useful properties which entanglement entropy enjoys generally.(See e.g. \cite{nielsen2000quantum}.)
We summarize some of them for later use.
\\1. If a composite system AB is in a pure state, then $S_A=S_B$.
\\2. If $\rho _{AB} = \rho_{A} \otimes \rho_{B} $,  then $  S_{AB} = S_A +S_B$.
\\3. For any subsystem $A$ and $B$, the following inequalities hold: 
%The subadditivity inequality and the triangle inequality 
\begin{gather}
S_{AB} \leq S_A +S_B  , \label{eq:sub ineq}     \\
S_{AB} \geq \mid S_A -S_B \mid . \label{eq:tri ineq}
\end{gather}
The first is the subadditivity inequality, and the second is the triangle inequality.

Because of translational invariance, $S_A$ and $S_B$ are independent of their positions, so,$\tfrac{\partial S_A}{\partial r}=0$ and $\tfrac{\partial  S_B}{\partial r}=0$.  
And the total system is in a pure state, so we have $S_{C} = S_{AB}$. Moreover, in the vacuum state, $\lim _{r\rightarrow \infty } \rho _{AB} = \rho_{A} \otimes \rho_{B} $  
because of the cluster decomposition principle \cite{footnote}. %\footnote{When d=1, this does not correct.}. 
So the property 2 suggests 
\begin{equation} 
\lim_{r\rightarrow \infty } S_C(r) = S_A +S_B . \label{eq:2-1}
\end{equation}
We apply (\ref{eq:sub ineq}) and (\ref{eq:tri ineq})to this system, then we obtain
\begin{equation} 
\mid S_A -S_B \mid \leq S_C(r) \leq S_A +S_B . \label{eq:2-2}
\end{equation}
Eqs. (\ref{eq:2-1}) and (\ref{eq:2-2}) show that $S_C$ (as a function of $r$) reaches its maximum value when $r\rightarrow \infty $.

\section{how to compute entanglement entropy} \label {review}
In this section we review the computational method developed by Bombelli et al \cite{Bombelli:1986rw}.
\subsection{Entanglement entropy of a collection of coupled harmonic oscillators}
We model the scalar field on $\mathbb{R}^d$ as a collection of coupled oscillators on a lattice of space points, 
labeled by capital Latin indices, the displacement at each point giving the value of the scalar field there.
In this case the Lagrangian can be given by
\begin{equation}
L=\dfrac{1}{2}G_{MN} \dot q ^M \dot q^N - \dfrac{1}{2} V_{MN} q^M q^N , \label{eq:3-1}
\end{equation}
where $q^M$ gives the displacement of the Mth oscillator and $\dot q^M$ its generalized velocity. 
The symmetric matrix $G_{MN}$ is positive definite, and therefore invertible, i.e, there exists the inverse matrix $G^{MN}$ such that
\begin{equation}
G^{MP} G_{PN} = \delta ^M_{~~ N} . \label{eq:3-2}
\end{equation}
The matrix $V_{MN}$ is also symmetric and positive definite.
Next, consider the positive definite symmetric matrix $W_{MN}$ defined by
\begin{equation}
W_{MA} G^{AB} W_{BN} = V_{MN} . \label{eq:3-3}
\end{equation}
The matrix $W$ is the square root of $V$ in the scalar product $G$.

Now consider a region $\Omega $ in $\mathbb{R}^d$.
The oscillators in this region will be specified by Greek letters, 
and those in the complement of $\Omega $ will be specified by lowercase Latin letters. 
%so that, e.g., the displacement of the $\alpha$th oscillator will be $q^\alpha $
We will use the following notation
\begin{alignat}{2}
W_{AB} = \begin{pmatrix} 
             W_{ab} & W_{a\beta}  \\
             W_{\alpha b} & W_{\alpha \beta} 
             \end{pmatrix} 
             \equiv   \begin{pmatrix} 
             A & B  \\
             B^T & C 
             \end{pmatrix}  &   ~~~~~~
 W^{AB} = \begin{pmatrix} 
             W^{ab} & W^{a\beta}  \\
             W^{\alpha b} & W^{\alpha \beta} \end{pmatrix}
              \equiv   \begin{pmatrix} 
             D & E  \\
             E^T & F 
             \end{pmatrix}             
, \label{eq:3-4}
\end{alignat}
where $W^{AB}$ is the inverse matrix of $W_{AB}$ ($W^{AB}$ is \textit{not} obtained by raising indices with $G^{AB}$).
So we have 
%Because $W_{AB}$ and $W^{AB}$ are positive definite matrices, $A,C,D$ and $F$ are
%positive definite matrices, and
\begin{alignat}{2}
 \begin{pmatrix} 
             1 & 0  \\
             0 & 1 
             \end{pmatrix} 
             =   \begin{pmatrix} 
             A & B  \\
             B^T & C 
             \end{pmatrix}  
   \begin{pmatrix} 
             D & E  \\
             E^T & F 
             \end{pmatrix}  
             =     \begin{pmatrix} 
            A D+BE^T & AE+BF  \\
             B^T D+CE^T & B^TE+ CF 
             \end{pmatrix}        
. \label{eq:3-5}
\end{alignat}
  
If we consider the information on the displacement of the oscillators inside $\Omega $ as unavailable, we can obtain a reduced density matrix 
$\rho_{red}$ for the outside $\Omega $, integrating out over $\mathbf{R}$ for each of the oscillators in the region $\Omega $, then we have
\begin{equation}
\rho_{red} ({q^a} , {q'^{b}}) =\int \prod_{\alpha } dq^{\alpha } \rho ({q^a, q^{\alpha }} , {q'^b,q^{\alpha}}) , \label{eq:3-6}
\end{equation}
where $\rho$ is a density matrix of the total system.

We can obtain the density matrix for the ground state by standard method, and it is a Gaussian density matrix.
Then,  $\rho_{red}$ is obtained by a Gaussian integral , and it is also a Gaussian density matrix.
The entanglement entropy, $S=-tr \rho_{red} \ln \rho_{red} $ , is given by \cite{Bombelli:1986rw}
\begin{gather}
 S=     \sum_n f(\lambda_n) ,  \label{eq:new3-1}  \\  
   f(\lambda) \equiv   \ln (\dfrac{1}{2} \lambda ^{1/2} ) + (1+\lambda  )^{1/2 } \ln [(1+\lambda  ^{-1})^{1/2 } + \lambda ^{-1/2} ]   ,
     \label{eq:3-7}
\end{gather}
where $\lambda_n$ are the eigenvalues of the matrix 
\begin{equation}
\Lambda ^a _{~b} = - W^{a\alpha} W_{\alpha b} =-(E B^T)^a_{~~b}.    \label{eq:3-8}
\end{equation}
It can be shown that all of $\lambda_n$ are nonnegative as follows.
From (\ref{eq:3-5}) we have
\begin{equation}
A \Lambda = -AEB^T = BFB^T .    \label{eq:3-9-1}
\end{equation}
It is easy to show that $A,C,D$ and $F$ are positive definite matrices when $W$ and $W^{-1}$ are positive definite matrices.
%$F$ is a positive definite matrix, 
Then $A\Lambda$ is a positive semi definite matrix as can be seen from (\ref{eq:3-9-1}).
So all eigenvalues of $\Lambda$ are nonnegative.
Finally, we can obtain the entanglement entropy by solving the eigenvalue problem of $\Lambda $.

%\begin{equation}
%W_{am} \Lambda ^m _{~b} = -W_{am} W^{m\alpha} W_{\alpha b} 
%=W_{a\gamma} W^{\gamma \alpha} W_{\alpha b} .    \label{eq:3-8}
%\end{equation}

\subsection{The continuum limit} \label{review 2}
Next, we apply the above formalism to  a massless free scalar field in (d+1) dimensional Minkowski spacetime.
We take the continuum limit in the above formalism. %of the formalism.
In this case the Lagrangian is given by
\begin{equation}
L= \int d^dx \dfrac{1}{2} [ \dot\phi ^2 - ( \nabla \phi  )^2 ].  \label{eq:3-14}
\end{equation}
Then the potential term becomes
%and we have
\begin{equation}
\dfrac{1}{2} V_{AB} q^A q^B  \rightarrow  \int d^d x \dfrac{1}{2} [ ( \nabla \phi  )^2  ] . \label{eq:3-8-1}
\end{equation}
The matrices $V,W$ and $W^{-1}$ are given in the momentum representation by,
\begin{gather}
V(x,y) = \int \dfrac{d^d k}{ (2\pi )^d } (k^2  ) e^{ik \cdot (x-y)}    \label{eq:3-9}  \\
W(x,y) = \int \dfrac{d^d k}{ (2\pi )^d } (k^2  )^{1/2}  e^{ik \cdot (x-y)}    \label{eq:3-10}  \\
W^{-1}(x,y) = \int \dfrac{d^d k}{ (2\pi )^d } (k^2  )^{-1/2}  e^{ik \cdot (x-y)}    \label{eq:3-11}
\end{gather}
From (\ref{eq:3-8}), the matrix $\Lambda$ is obtained as a sum over the oscillators in the region $\Omega $,
\begin{equation}
\Lambda (x,y) = - \int_{\Omega } d^d z W^{-1} (x,z) W(z,y) .   \label{eq:3-12}
\end{equation}
We now have to solve the eigenvalue equation
\begin{equation}
\int_{\Omega^c } d^d y \Lambda (x,y) f(y) = \lambda f(x) ,   \label{eq:3-13}
\end{equation}
where $\Omega^c$ is the complementary set of $\Omega$,
and then we use the eigenvalues in the expression for the entropy (\ref{eq:new3-1}).

 %\begin{figure}
 %\includegraphics[width=11zw,angle=270]{paperfigw3.eps}%
 %\caption{\label{fig W}}
 %\end{figure}

%\begin{figure}[t]
 %\includegraphics[width=30zw,angle=270,clip]{paperfig0-2-2.eps}
 %\caption{\label{fig0-2}}
 %\end{figure}

% \begin{figure*}
 %\includegraphics[width=30zw,angle=270,clip]{wcolor-2.eps}%
 %\caption{\label{fig W}}
% \end{figure*}

%\begin{figure}
% \includegraphics[width=25zw,angle=270,clip]{paperfiglambda-2.eps}%
% \caption{\label{fig lambda}}
 %\end{figure}

%\begin{figure}
% \includegraphics[width=30zw,angle=270,clip]{paperfiglambda0-2.eps}%
% \caption{\label{fig lambda0}}
% \end{figure}

%\begin{figure}
% \includegraphics[width=30zw,angle=270,clip]{deltalambda-2.eps}%
 %\caption{\label{deltalambda}}
% \end{figure}

%\begin{figure}
% \includegraphics[width=30zw,angle=270,clip]{deltalambda1-2.eps}%
% \caption{\label{fig deltalambda1}}
 %\end{figure}

%\begin{figure*}
% \includegraphics[width=10zw,angle=270,clip]{deltalambdad-2.eps}%
% \caption{\label{fig deltalambda d}}
% \end{figure*}

\section{entanglement entropy of two disjoint regions in a $d+1$ dimensional massless free scalar field} \label{flat}
\begin{figure}
 \includegraphics[width=6.5cm,angle=270,clip]{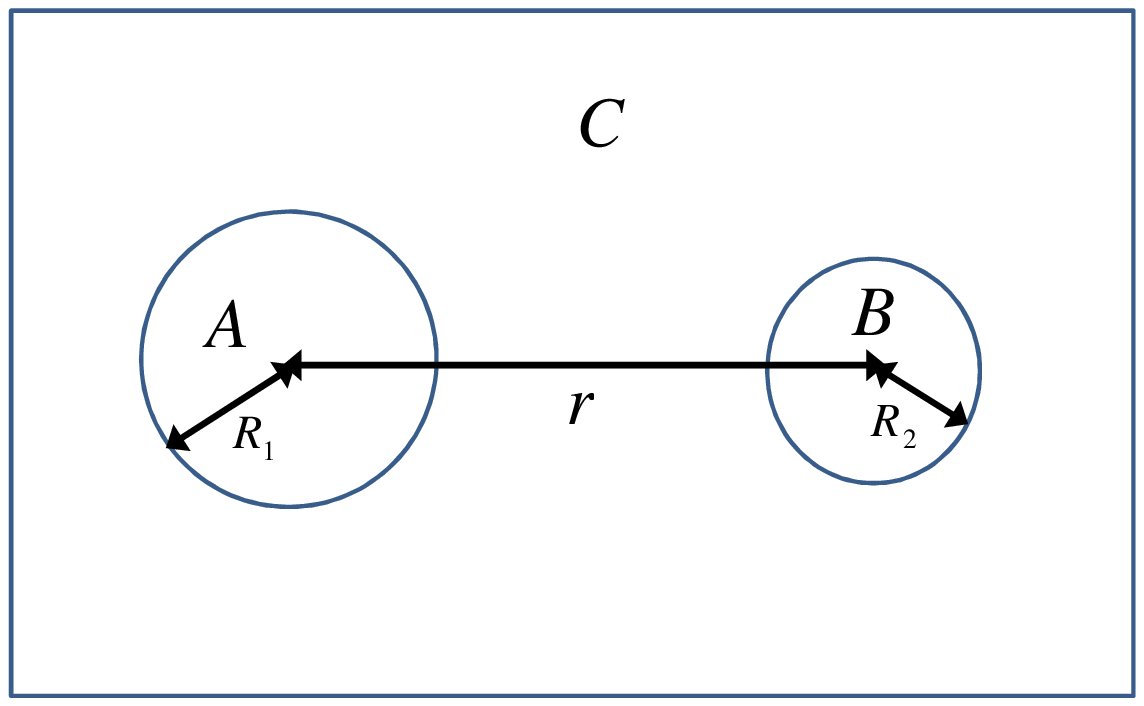}%
 \caption{Two spheres $A$ and $B$, and the outside region $C$. }
 \label{threeregions}
 \end{figure}

We consider two spheres $A$ and $B$ whose radii are $R_1$ and $R_2$, and define the outside region as $C$. (See Fig \ref{threeregions}.)
We derive the $r(\gg R_1,R_2)$ dependence of $S_C(r,R_1,R_2)$ by using the formalism of the preceding section.
%Next we consider the shapes of $A$ and $B$.
(In the later analysis we do not use the shapes of $A$ and $B$, so 
all analysis in this section holds for $A$ and $B$ which have arbitrary shapes.
In this case $R_1$ and $R_2$ are the characteristic sizes of $A$ and $B$.)

We consider $S_{AB} (r,R_1,R_2)$ because $S_C=S_{AB}$ in a pure state and the $r$ dependence of $S_{AB}$ is clearer than that of $S_{C}$ in the calculation.
In this case the region $\Omega$ is $C$. %$\Omega =C$. 

We obtain the $r$ dependence of $S_{AB}$ by following three steps:

(1) We obtain the $r$ dependence of $\Lambda$ by using the $\|x-y\| $ dependences of $W(x,y)$ and $W^{-1}(x,y)$.
We decompose $\Lambda$ into the non-perturbative part and the perturbative part as $\Lambda=\Lambda^{0} + \delta \Lambda $,
where $\Lambda^{0} \equiv \lim_{r\rightarrow \infty} \Lambda$.

(2) We obtain $\lambda_m (r)$ which are the eigenvalues of $\Lambda$ by perturbation theory.
This is almost similar to the time-independent perturbation theory 
in quantum mechanics in the presence of degeneracy. %with degeneracy of quantum mechanics.
We can regard $\Lambda$ as Hamiltonian. Note that $\Lambda$ is \textit{not} a symmetric matrix.
So we must slightly modify the perturbation theory in quantum mechanics.

(3) In Step (2), we had $\lambda_m (r)$ as $\lambda_m(r) = \lambda_m^{0} +\delta \lambda_m(r)  $, where $\lambda_m^{0}$ are  the eigenvalues of $\Lambda^0$. 
We substitute these $\lambda_m (r)$ into (\ref{eq:3-7}), 
then we obtain $S_{AB}(r,R_1,R_2)$.

First we examine the $\|x-y\| $ dependences of $W(x,y)$ and $W^{-1}(x,y)$.
Generally entanglement entropy has UV divergence as discussed in \cite{Bombelli:1986rw}.
So we use a momentum cutoff $l^{-1}$ in integrals (\ref{eq:3-9})-(\ref{eq:3-11}), though these integrals are well defined as Fourier transforms of distributions. 
(The other regularization methods are discussed in \cite{Bombelli:1986rw}.)
When $d\geq 2$ and $ l / \|x-y\| \rightarrow 0 $ , $W(x,y)$ and $W^{-1}(x,y)$ are
\begin{equation}
W(x,y) = \dfrac{A_d}{\|x-y\| ^{d+1}} ~, ~~~~W^{-1}(x,y) =  \dfrac{B_d}{\|x-y\| ^{d-1}} ,~~~A_d ,B_d \in \mathbb{R}
\label{eq:4-8}
\end {equation}
where $A_d$ and $B_d$ are nonzero dimensionless constants (see Appendix A).
We \textit{cannot} obtain ($\ref{eq:4-8}$) by only using a dimensional analysis because 
 $ l / \|x-y\| $ is dimensionless.
 Indeed $V(x,y) \rightarrow 0$ when $ l / \|x-y\| \rightarrow 0 $, i.e. $V(x,y)$ is zero when $\|x-y\| $ is finite.
%$\delta(\|x-y\|)$ is possible .
%In a momentum cutoff regularization this corresponds to $\lim_{l / \|x-y\| \rightarrow 0 } W or W^{-1} =0.$  
%Indeed $V(x,y)$ is a derivative of $\delta^d(x-y)$ so $V(x,y) \rightarrow 0$ when $ l / \|x-y\| \rightarrow 0 $.
On the other hand $W$ and $W^{-1}$ have nonzero value for $\|x-y\| > 0$ 
because they are kernels of integral operators of nonlocal interaction (i.e Fourier transformations of $(\sqrt{k^2})^{\pm 1}$) .
In Appendix A we explicitly show that $W$ and $W^{-1}$ have nonzero value for $\|x-y\| > 0$ and Eq. (\ref{eq:4-8}) holds. 
%Explicit calculation is in the Appendix.

%Next, we obtain the $r$ dependence of $\Lambda$ by using (\ref{eq:4-8}) .
%For this purpose we divide $\mathbb{R}^d$ into $A,B,\bar{A}$ and $\bar{B}$, see Fig \ref{tworegion}.
%We define $\mathbb{R}_A^d (\mathbb{R}_B^d)$ as the half space which include $A(B)$, and its boundary is at the middle point between $A$ and $B$.
%Then $\bar{A} = \mathbb{R}_A^d - A$ , $\bar{B} = \mathbb{R}_B^d - B$ and $\mathbb{R}^d =A+\bar{A} +B+\bar{B}$.

\begin{figure}
 \includegraphics[width=6.5cm,angle=270,clip]{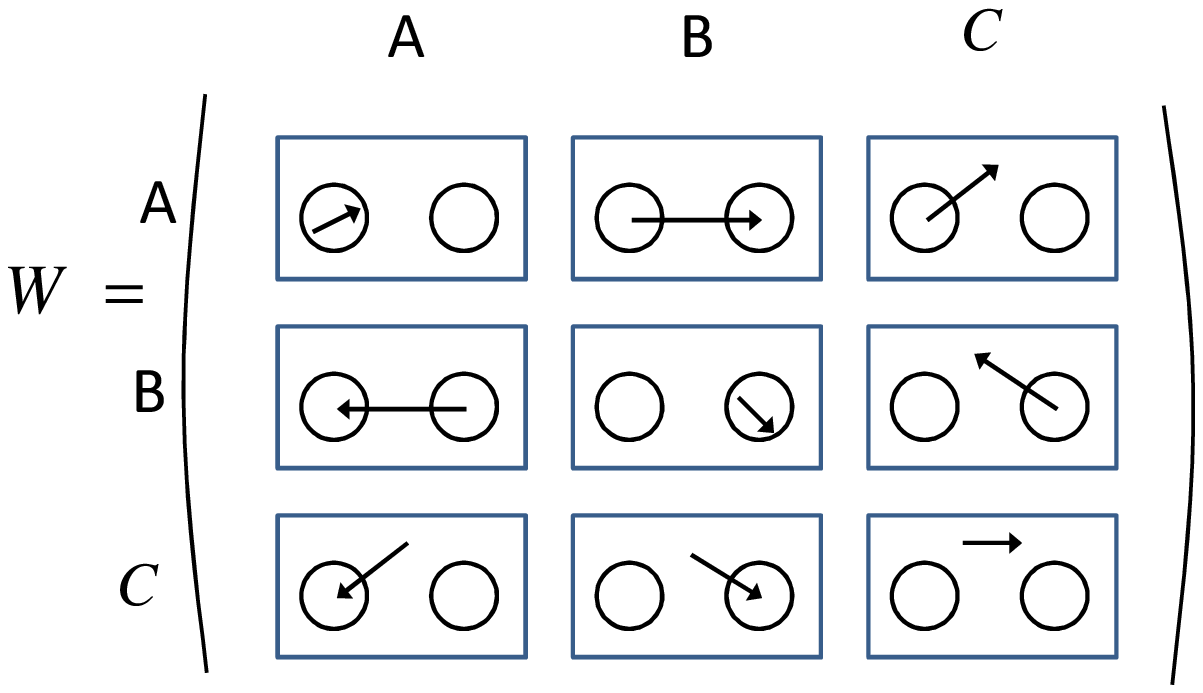}%
 \caption{The matrix elements of W. The lines denote the matrix elements $W(x,y)$  in (\ref{eq:4-8}). 
An initial point and an end point of an arrow denote a row and a column respectively.
We can obtain products of matrices by connecting arrows and integrating joint points on regions where the joint points exist. Instead of solid lines we use dotted lines for $W^{-1}$.   }
 \label{Wfig}
 \end{figure}

\begin{figure}
 \includegraphics[width=6.5cm,angle=270,clip]{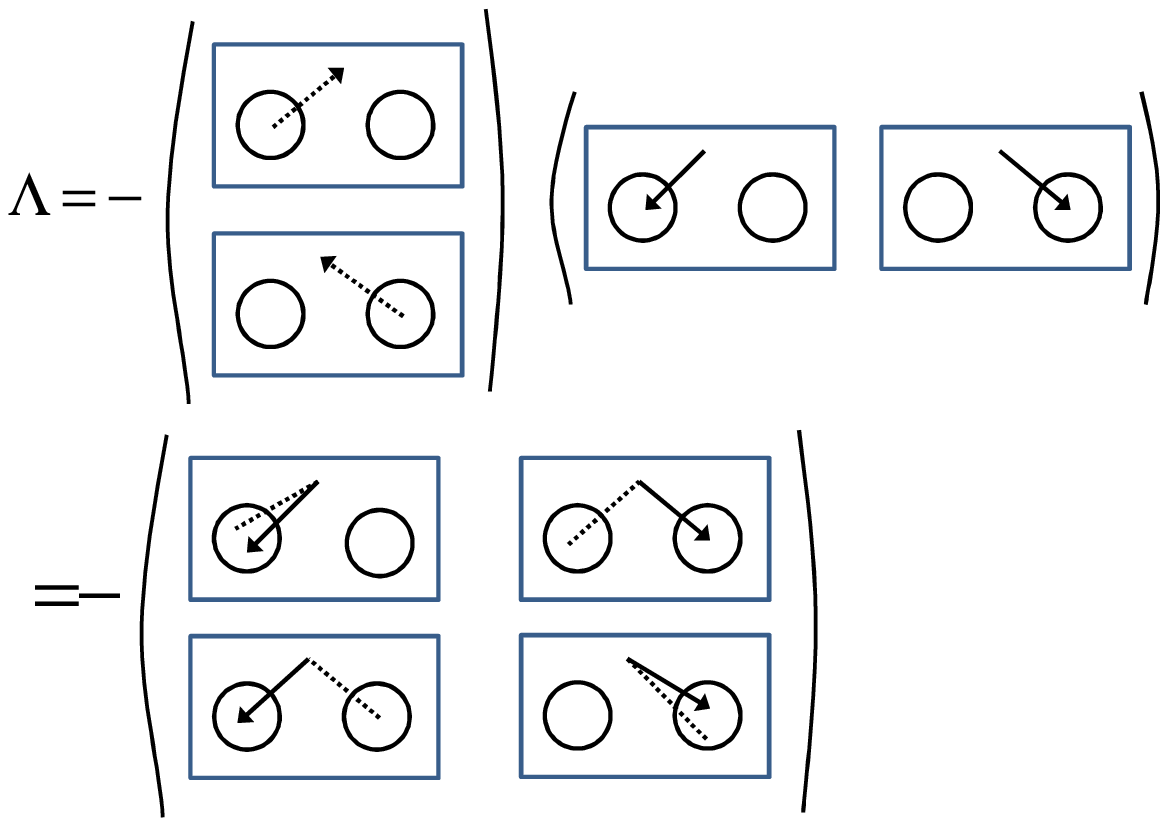}%
 \caption{The diagrammatic calculation of $\Lambda$ in (\ref{eq:new4-1}).}
 \label{lambda}
 \end{figure}

\begin{figure}
 \includegraphics[width=6.5cm,angle=270,clip]{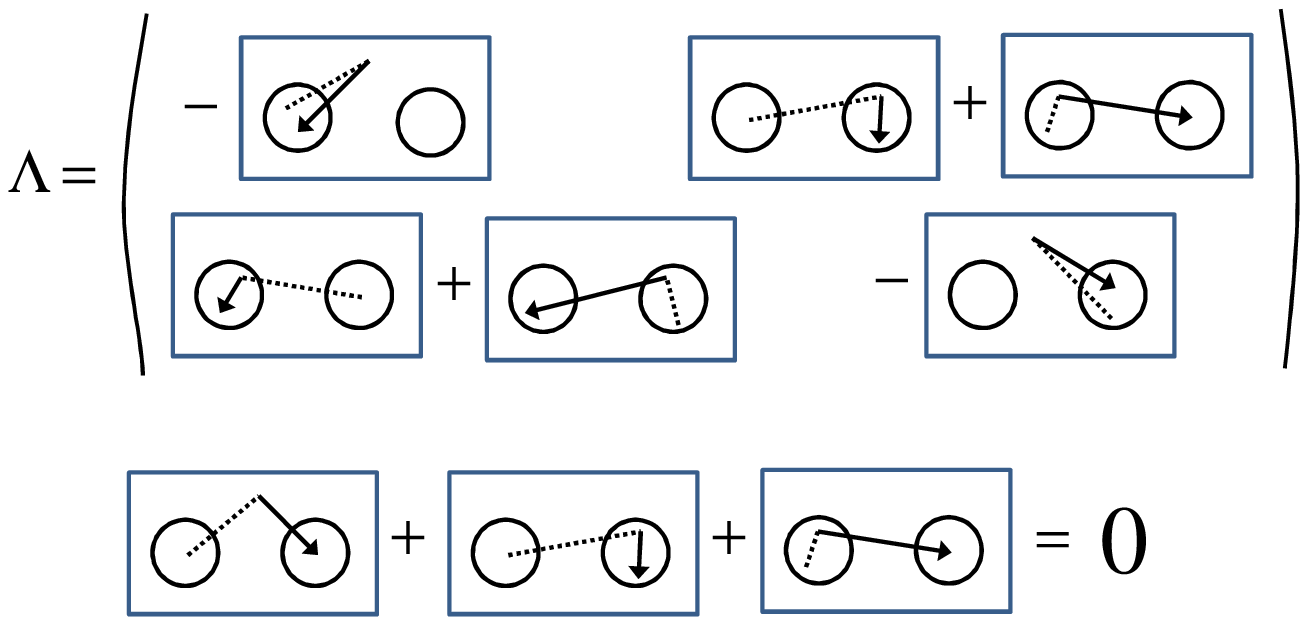}%
 \caption{The diagrammatic representations of $\Lambda$ in (\ref{eq:new4-3}) and the identity in (\ref{eq:new4-2}). }
 \label{lambda2}
 \end{figure}

\begin{figure}
 \includegraphics[width=6.5cm,angle=270,clip]{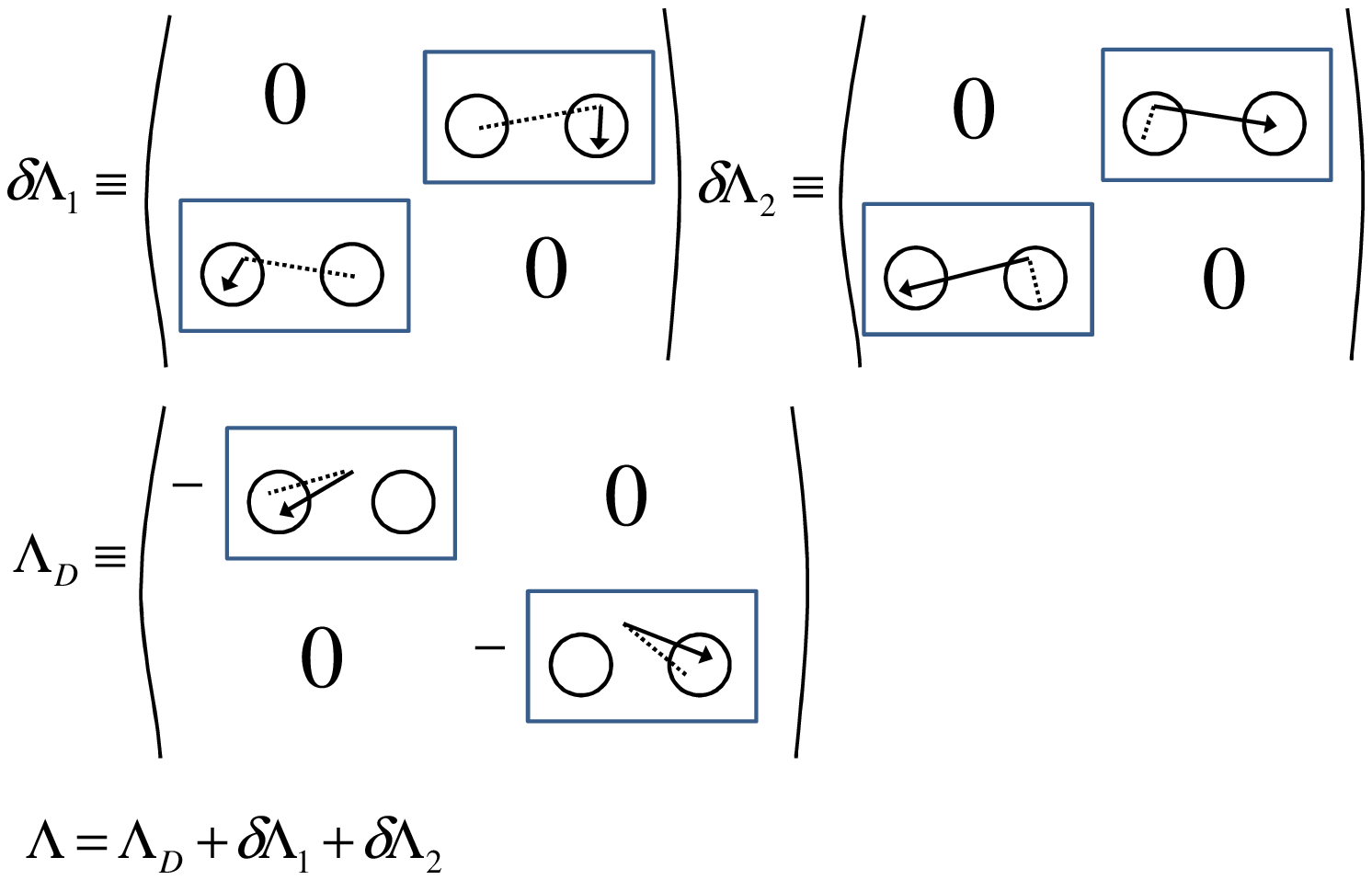}%
 \caption{The diagrammatic representations of $\delta \Lambda_1$, $\delta \Lambda_2$ and $ \Lambda_D$.}
 \label{dellambda}
 \end{figure}

\begin{figure}
 \includegraphics[width=6.5cm,angle=270,clip]{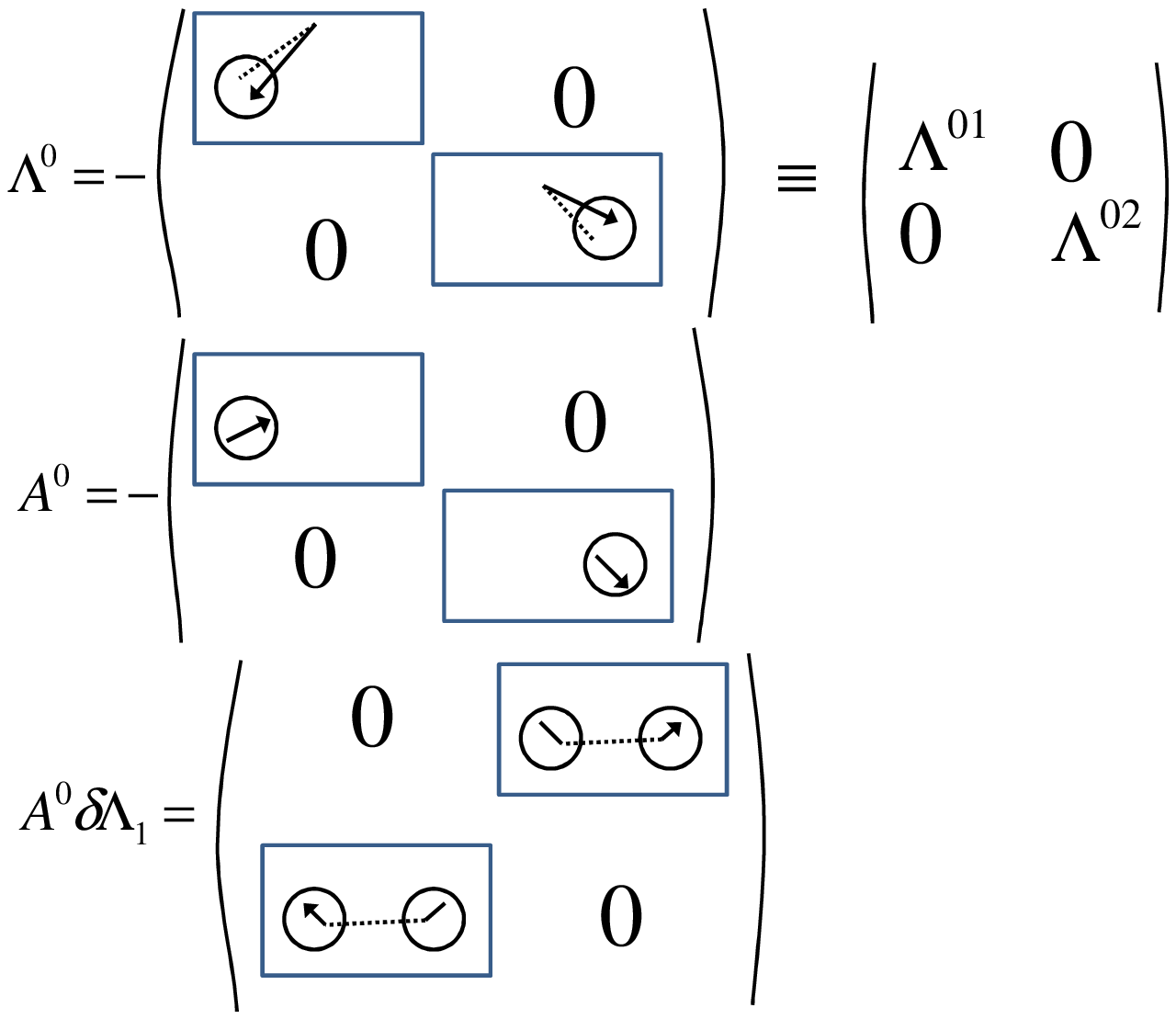}%
 \caption{The diagrammatic representations of $\Lambda^0$, $A^0$ and $ A^0 \delta \Lambda_1$.}
 \label{lambda0}
 \end{figure}

\begin{figure}
 \includegraphics[width=6.5cm,angle=270,clip]{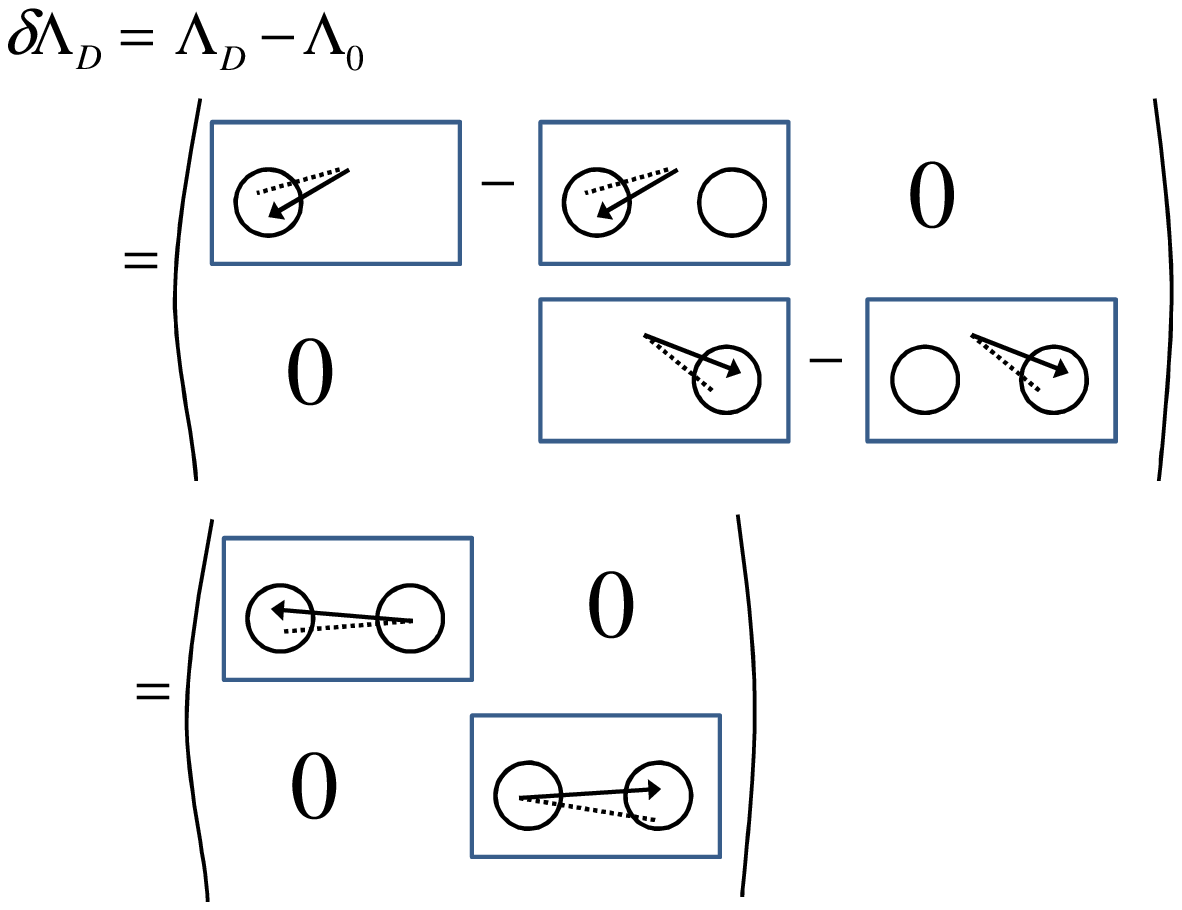}%
 \caption{The diagrammatic calculation of $\delta \Lambda_D$ in (\ref{eq:new4-12}).}
 \label{dellambdad}
 \end{figure}

Next, we obtain the $r$ dependence of $\Lambda$ by using (\ref{eq:4-8}).
We represent the matrix elements of $W(W^{-1})$ diagrammatically in Fig \ref{Wfig}.
%We change solid lines to dotted lines for $W^{-1}$.  
Instead of solid lines we use dotted lines for $W^{-1}$.   
The lines denote the matrix elements $W(x,y)$(or $W^{-1}(x,y)$)  in (\ref{eq:4-8}). %\textit{propagators} in (\ref{eq:4-8}).
An initial point and an end point of an arrow denote a row and a column respectively.
We can obtain products of matrices by connecting arrows and integrating joint points on regions where the joint points exist.
We label coordinates in $A, B $ and $C$ as $x_a, x_b$ and $x_c$.
Then, from Fig \ref{lambda} we obtain $\Lambda = -E B^T$  as
\begin{equation}
\begin{split}
  \Lambda 
 &=  \begin{pmatrix}   \Lambda (x_a,y_a)  & \Lambda (x_a,y_b)  \\
                                \Lambda  (x_b,y_a) & \Lambda (x_b,y_b)  \end{pmatrix} 
= - \begin{pmatrix}   W^{-1} (x_a,z_c)    \\
                               W^{-1} (x_b,z_c)   \end{pmatrix} 
       \begin{pmatrix}   W (z_c,y_a)    &
                               W(z_c,y_b)   \end{pmatrix}    \\                    
&= - \begin{pmatrix}  \int_{C} d^{d} z_c W^{-1} (x_a,z_c) W (z_c,y_a) & \int_{C} d^{d} z_c W^{-1} (x_a,z_c)  W(z_c,y_b) \\
                             \int_{C} d^{d} z_c  W^{-1} (x_b,z_c)  W (z_c,y_a)  &  \int_{C} d^{d} z_c  W^{-1} (x_b,z_c)  W(z_c,y_b)   \end{pmatrix} .
\label{eq:new4-1}
\end{split}
\end{equation}
To make the $r$ dependence of the non-diagonal elements of $\Lambda$ clear,
we use the following identity,
\begin{equation}
\int_{A+B+C} d^d z W^{-1} (x_a,z) W(z,y_b) = \delta (x_a -y_b) =0 .
\label{eq:new4-2}
\end{equation} 
We represent this identity diagrammatically in Fig \ref{lambda2}.
From (\ref{eq:new4-1}) and (\ref{eq:new4-2}) we obtain (see Fig \ref{lambda2})
\begin{equation}
\begin{split}
&\Lambda (x_a,y_b) = \int_{A} d^{d} z_a W^{-1} (x_a,z_a)  W(z_a,y_b) + \int_{B} d^{d} z_b W^{-1} (x_a,z_b)  W(z_b,y_b)   \\
&\Lambda  (x_b,y_a) = \int_{A} d^{d} z_a  W^{-1} (x_b,z_a)  W(z_a,y_a) + \int_{B} d^{d} z_b  W^{-1} (x_b,z_b)  W(z_b,y_a)  .
\label{eq:new4-3}
\end{split}
\end{equation}  
Note that from (\ref{eq:4-8}) $W(x,y)$ and $W^{-1}(x,y)$ have the different $\| x-y \|$ dependence.
So, from (\ref{eq:new4-1}) and (\ref{eq:new4-3}) we decompose $\Lambda$ as
\begin{equation}
\Lambda = \Lambda_D + \delta \Lambda_1 + \delta \Lambda_2  
\label{eq:new4-4}
\end{equation}  
where we define (see Fig \ref{dellambda})
\begin{align}
& \Lambda_D \equiv \begin{pmatrix}   \Lambda (x_a,y_a)  & 0  \\
                                0 & \Lambda (x_b,y_b)  \end{pmatrix} , \label{eq:new4-5} \\
&\delta \Lambda_1 \equiv \begin{pmatrix}   0  & \int_{B} d^{d} z_b W^{-1} (x_a,z_b)  W(z_b,y_b)   \\
                                \int_{A} d^{d} z_a  W^{-1} (x_b,z_a)  W(z_a,y_a) & 0  \end{pmatrix} , \label{eq:new4-6} \\
& \delta \Lambda_2   \equiv \begin{pmatrix}   0  &  \int_{A} d^{d} z_a W^{-1} (x_a,z_a)  W(z_a,y_b)    \\
                               \int_{B} d^{d} z_b  W^{-1} (x_b,z_b)  W(z_b,y_a)   & 0  \end{pmatrix}  \label{eq:new4-7}  .                             
\end{align}  
We approximate $W(x_a,y_b)\approx \tfrac{A_d}{r^{d+1}}$ and $ W^{-1}(x_a,y_b)\approx \tfrac{B_d}{r^{d-1}}$ 
because $r\gg R_1,R_2$.
Then we have
\begin{align}
&\delta \Lambda_1 \approx \dfrac{B_d}{r^{d-1}} \begin{pmatrix}   0  & \int_{B} d^{d} z_b   W(z_b,y_b)   \\
                                \int_{A} d^{d} z_a    W(z_a,y_a) & 0  \end{pmatrix} , \label{eq:new4-8} \\
& \delta \Lambda_2   \approx \dfrac{A_d}{r^{d+1}} \begin{pmatrix}   0  &  \int_{A} d^{d} z_a W^{-1} (x_a,z_a)     \\
                               \int_{B} d^{d} z_b  W^{-1} (x_b,z_b)   & 0  \end{pmatrix}  \label{eq:new4-9}  .                             
\end{align}

Next we consider the non-perturbative part $\Lambda^0 = \lim_{r\rightarrow \infty} \Lambda$.
From (\ref{eq:new4-8}) and (\ref{eq:new4-9}) we can see that 
$\delta \Lambda_1$ and $\delta \Lambda_2$ become $0$ when $r\rightarrow \infty$.
Note that the integral region of the integral in $\Lambda(x_a,y_a)$ ($\Lambda(x_b,y_b)$) 
 become $A^c \equiv \mathbb{R}^d -A$ ($B^c \equiv \mathbb{R}^d -B$) when $r\rightarrow \infty$, then we obtain (see Fig \ref{lambda0}) %in Fig \ref{lambda0}.
\begin{equation}
  \Lambda^0 = - \begin{pmatrix}  \int_{A^c} d^{d} z W^{-1} (x_a,z) W (z,y_a) & 0  \\
                            0  &  \int_{B^c} d^{d} z  W^{-1} (x_b,z)  W(z,y_b)   \end{pmatrix} .
\label{eq:new4-10}
\end{equation}
%By decomposing $\Lambda_D$ in (\ref{eq:new4-4})
From %(\ref{eq:new4-4}) and 
(\ref{eq:new4-10}) we rewrite (\ref{eq:new4-4}) as follows,
\begin{equation}
\Lambda = \Lambda^0 + \delta \Lambda_1 + \delta \Lambda_2 + \delta \Lambda_D  ,
\label{eq:new4-11}
\end{equation}  
where we define (see Fig \ref{dellambdad})
\begin{equation}
 \delta \Lambda_D \equiv \Lambda_D - \Lambda^0 
= \begin{pmatrix}  \int_{B} d^{d} z_b W^{-1} (x_a,z_b) W (z_b,y_a) & 0  \\
                            0  &  \int_{A} d^{d} z_a  W^{-1} (x_b,z_a)  W(z_a,y_b)   \end{pmatrix} .
\label{eq:new4-12}
\end{equation}  
We use the same approximation as %was 
we used in (\ref{eq:new4-8}) and (\ref{eq:new4-9}),
then we obtain
\begin{equation}
 \delta \Lambda_D 
\approx   \dfrac{A_d B_d}{r^{2d}}  \begin{pmatrix}  \int_{B} d^{d} z_b  & 0  \\
                            0  &  \int_{A} d^{d} z_a    \end{pmatrix} .
\label{eq:new4-13}
\end{equation}  
When we perform the perturbative calculation to obtain $\lambda_m (r)$ which is the eigenvalues of $\Lambda$,
from (\ref{eq:new4-8}), (\ref{eq:new4-9}), (\ref{eq:new4-11}) and (\ref{eq:new4-13})  
we can neglect $\delta \Lambda_D$ because it is higher order than $\delta \Lambda_1$ and $\delta \Lambda_2$ with respect to $1/r$.
And we can neglect $\delta \Lambda_2$ because its nonzero matrix elements are in the same position as   
$\delta \Lambda_1$ and $\delta \Lambda_2$ is higher order than $\delta \Lambda_1$ with respect to $1/r$.
%This approximation will be justified in later perturbative calculations.

Because $\Lambda$ is not a symmetric matrix, in the later perturbative calculation 
we need $A^0 \delta \Lambda_1$ where $A^0$ is defined as (see Fig \ref{lambda0})
\begin{equation}
 A^0 \equiv \lim_{r\rightarrow \infty} A = \begin{pmatrix}  W(x_a,y_a) & 0  \\
                            0  &  W(x_b,y_b)   \end{pmatrix} .
\label{eq:new4-14}
\end{equation} 
From (\ref{eq:new4-6}), (\ref{eq:new4-8}) and (\ref{eq:new4-14}) we obtain (see Fig \ref{lambda0})
{\small \begin{equation}
\begin{split}
 A^0 \delta \Lambda_1 
&=\begin{pmatrix}  0  & \int_{A} d^d z_a \int_{B} d^d z_b W(x_a, z_a)W^{-1}(z_a,z_b)W(z_b, y_b )     \\
                      \int_{B} d^d z_b \int_{A} d^d z_a  W(x_b, z_b ) W^{-1} (z_b,z_a) W(z_a, y_a)    &  0    \end{pmatrix}  \\
& \approx 
 \dfrac{B_d}{r^{d-1}}  
\begin{pmatrix}  0  & \int_{A} d^d z_a W(x_a, z_a)  \int_{B} d^d z_b W(z_b, y_b )     \\
                      \int_{B} d^d z_b W(x_b, z_b ) \int_{A} d^d z_a W(z_a, y_a)    &  0    \end{pmatrix} .
\label{eq:new4-15}
\end{split}
\end{equation} }
We have finished the first step.

Next, we calculate $\lambda_m(r)$ by perturbation theory.
This is almost similar to the time-independent perturbation theory with degeneracy of quantum mechanics.
The only difference is that $\Lambda$ is \textit{not} a symmetric matrix and $A \Lambda$ is a symmetric matrix.

We can approximate $\Lambda \approx \Lambda^0 +\delta \Lambda_1 $ and regard $\delta \Lambda_1 $ as the perturbative part.
Then, from (\ref{eq:new4-8}) we expand  $\lambda_m(r)$ with respect to $1/r^{d-1}$.
We expand $\lambda_m$ around $\lambda_m^0\equiv  \lambda_m (r=\infty)$,
\begin{equation}
\lambda_m = \lambda_m^0 + \delta \lambda_m^{1} +\delta \lambda_m^{2} ,  
\label{eq:4-12}
\end{equation}
where $\delta \lambda_m^{1}$ and $\delta \lambda_m^{2}$ are the first and the second order perturbations.

 Next we substitute (\ref{eq:4-12}) into (\ref{eq:3-7}),
\begin{equation}
\begin{split}
S_{AB}(r,R_1,R_2) & = \sum_m f( \lambda_m) \\
&= S_A(R_1) + S_B(R_2) +\sum_m 
\left[  \delta \lambda_m \left. \dfrac{df}{d\lambda_m} \right|_{\lambda_m = \lambda_m^0} 
+\dfrac{1}{2} (\delta \lambda_m )^2 \left. \dfrac{d^2f}{d\lambda_m^2} \right|_{\lambda_m = \lambda_m^0} \right] ,  
\label{eq:4-13}
\end{split}
\end{equation}
where $\delta \lambda_m \equiv  \delta \lambda_m^{1} +\delta \lambda_m^{2} $.
We will show that the first order perturbation in (\ref{eq:4-13}) (i.e $\sum_m \delta \lambda_m^{1} \left. \dfrac{df}{d\lambda_m} \right|_{\lambda_m = \lambda_m^0} $) 
is zero, %cancel each other, 
so we must calculate the second order perturbations.

We label the $\lambda_m^0$'s as $\lambda_m^0 >\lambda_n^0  ~\textrm{when} ~ m>n $. 
And we define the eigenvectors of $\Lambda^0$,
\begin{equation}
f_{m1\alpha }^0 =
\begin{pmatrix} f_{m1\alpha }^0 (x_a)  \\
0   \end{pmatrix} ,
f_{n2\beta  }^0 =
\begin{pmatrix} 0 \\ f_{n2\beta  }^0 (x_b)  \end{pmatrix} ,
\Lambda^0 f_{m1\alpha }^0 = \lambda_m^0 f_{m1\alpha }^0, 
\Lambda^0 f_{m2\beta  }^0 = \lambda_m^0 f_{m2\beta  }^0, 
\label{eq:4-14}
\end{equation} 
where $\alpha=1,\cdots, M_m $ and $\beta=1,\cdots,N_n $ are the labels of the degeneracy.
And we normalize $f_{mi\alpha }^0 ~~(i=1,2)$ as follow,
\begin{equation}
f_{mi\alpha }^{0T} A^0 f_{nj\beta  }^0= \delta_{mn} \delta_{ij} \delta_{\alpha \beta }  .
\label{eq:4-15}
\end{equation} 
This normalization is always possible because $A^0$ is a positive definite symmetric matrix.
For general $R_1$ and $R_2$, $\Lambda^{01}$ and $\Lambda^{02}$ have different eigenvalues,
 so there are two groups of $\lambda_m^0$; one is the group of the common eigenvalues of $\Lambda^{01}$ and $\Lambda^{02}$,
 the other is not. %the class of the eigenvalues of $\Lambda^{01}$ and not of $\Lambda^{02}$.
We will see that $\delta \lambda_m^1$ of the latter group are zero.
We expand $f_{m\gamma  }$ which is the eigenvector of $\Lambda$ in the following way, %as follow, 
\begin{equation}
f_{m\gamma  } = \sum_\alpha a_{\gamma \alpha }  f_{m1\alpha }^0  +  \sum_\beta  b_{\gamma \beta  }  f_{m2\beta  }^0  
                      + f_{m\gamma  }^1 + f_{m\gamma  }^2    \equiv  \xi_{m\gamma }^0  + f_{m\gamma  }^1 + f_{m\gamma  }^2                      
\label{eq:4-16}
\end{equation} 
where $f_{m\gamma  }^1$ and $f_{m\gamma  }^2$ are the first and the second order perturbations.
Note that when $\lambda_m^0$ is an eigenvalue of $\Lambda^{01}$ ($\Lambda^{02}$) and is not an eigenvalue of $\Lambda^{02}$ ($\Lambda^{01}$), then  
%common eigenvalue of $\Lambda^{01}$ and $\Lambda^{02}$ , 
the coefficients $b_{\gamma \beta  }$  ($a_{\gamma \alpha }$) are  zero ; because the zeroth order eigenvectors $f_{m2\beta  }^0$ ($f_{m1\alpha }^0$) do not exist.
So either the coefficients $a_{\gamma \alpha }$ or $b_{\gamma \beta  }$ are zero when $\lambda_m^0$ is not a common eigenvalue of $\Lambda^{01}$ and $\Lambda^{02}$.
We substitute (\ref{eq:4-16}) into the eigenvalue equation (we approximate $\Lambda \approx  \Lambda^0 +\delta \Lambda_1 $) , then we have
\begin{equation}
(\Lambda^0 + \delta \Lambda_1 ) f_{m\gamma  } = (\lambda_m^0 + \delta \lambda_{m\gamma }^1 +\delta \lambda_{m\gamma }^2  )   f_{m\gamma  }  .                
\label{eq:4-17}
\end{equation} 
We obtain equations of the first and the second order perturbation. 
\begin{align}
& \Lambda^0  f_{m\gamma}^1 + \delta \Lambda_1  \xi_{m\gamma }^0 = \lambda_m^0 f_{m\gamma  }^1  + \delta \lambda_{m\gamma }^1 \xi_{m\gamma }^0 ,  \label{eq:4-18}  \\   
& \Lambda^0  f_{m\gamma}^2 + \delta \Lambda_1 f_{m\gamma }^1 = \lambda_m^0 f_{m\gamma}^2+ \delta \lambda_{m\gamma }^1 f_{m\gamma  }^1 +\delta \lambda_{m\gamma }^2
 \xi_{m\gamma  }^0 .                  
\label{eq:4-19}
\end{align} 
We multiply (\ref{eq:4-18}) by $f_{mj \gamma ' }^{0T} A^0$ from the left .
The first term of the left hand side of (\ref{eq:4-18}) cancel the first term of the right hand side of (\ref{eq:4-18}) 
because $A^0 \Lambda^0$ is a symmetric matrix, then we obtain
\begin{align}
\sum_\alpha a_{\gamma \alpha } V_{m\gamma ' m \alpha }^{j1}  +  \sum_\beta  b_{\gamma \beta  }  V_{m\gamma ' m\beta }^{j2} 
=  \delta \lambda_{m\gamma }^1 (a_{\gamma \gamma '  } \delta ^{j1} +b_{\gamma \gamma '  } \delta ^{j2}  ) ,             
\label{eq:4-20}
\end{align} 
where
\begin{align}
V_{m\alpha n\beta } ^{ij} \equiv   f_{mi\alpha }^{0T} A^0  \delta \Lambda_1 f_{nj\beta  }^0 .
\label{eq:4-21}
\end{align} 
From (\ref{eq:new4-15}) we obtain  $V_{m\alpha n\beta } ^{11} = V_{m\alpha n\beta } ^{22} =0  $ and
\begin{align}
\begin{split}
V_{m\alpha n\beta } ^{12}  &= 
\begin{pmatrix} f_{m1\alpha }^0 (x_a)  &
0   \end{pmatrix} 
 \begin{pmatrix} A^0 \delta \Lambda_1 (x_a,y_a)  & A^0\delta \Lambda_1 (x_a,y_b)  \\
                               A^0 \delta \Lambda_1 (x_b,y_a) & A^0 \delta \Lambda_1 (x_b,y_b)  \end{pmatrix} 
\begin{pmatrix} 0 \\ f_{n2\beta  }^0 (y_b)  \end{pmatrix} \\
&=  \dfrac{B_d}{r^{d-1}}  
\int_{A} d^dx_a \int_{A} d^d z_a W(x_a, z_a) f_{m1\alpha }^0 (x_a) \int_{B} d^dy_b \int_{B} d^d z_b W(y_b, z_b ) f_{n2\beta  }^0 (y_b) 
\equiv \dfrac{B_d}{r^{d-1}} C_{m\alpha n\beta } 
\label{eq:4-22}
\end{split}
\end{align} 
and $V_{m\alpha n\beta } ^{12}  = V_{  n\beta m\alpha } ^{21} $.
We define an $M_m \times N_n$ matrix $C_{mn}$ as $(C_{mn})_{\alpha \beta } =C_{m\alpha n\beta }$ and write (\ref{eq:4-20}) as follows,
\begin{equation}
\dfrac{B_d}{r^{d-1}} 
\begin{pmatrix} 0  & C_{mm}  \\
                      C_{mm}^T    &   0   \end{pmatrix} 
\begin{pmatrix} \mathbf{a}_{\gamma } \\
                     \mathbf{b}_{\gamma }  \end{pmatrix}                       
= \delta \lambda_{m\gamma }^1
 \begin{pmatrix} \mathbf{a}_{\gamma } \\
                     \mathbf{b}_{\gamma }  \end{pmatrix}                           
\label{eq:4-23}
\end{equation}
where $(\mathbf{a}_{\gamma } )_{\alpha } =a_{\gamma \alpha } $ and $(\mathbf{b}_{\gamma } )_{\beta  } =b_{\gamma \beta  } $.
From (\ref{eq:4-23}), if $\lambda_m^0$ is not a common eigenvalue of $\Lambda^{01}$ and $\Lambda^{02}$,
$\delta \lambda_{m\gamma }^1 $ is zero; 
because either $a_{\gamma \alpha }$ or $b_{\gamma \beta  }$ are zero when $\lambda_m^0$ is not a common eigenvalue of $\Lambda^{01}$ and $\Lambda^{02}$. 
%We assume that $M_m > N_m$  without loss of generality. 
We first consider the case that $M_m \geq  N_m$.
In this case we obtain the following eigenvalue equation \cite{footnote2}.
\begin{equation}
\begin{split}
\det 
\begin{vmatrix} x 1_{M_m \times M_m}  &- C_{mm}  \\
                     - C_{mm}^T    &   x 1_{N_m \times N_m}   \end{vmatrix} 
&= \det (x 1_{M_m \times M_m} ) \det ( x 1_{N_m \times N_m} -x^{-1}  C_{mm}^T C_{mm} )  \\ 
& = x^{M_m -N_m} \det (x^2 1_{N_m \times N_m} -  C_{mm}^T C_{mm} ) =0  .              
\label{eq:4-24}
\end{split}
\end{equation}
We define the eigenvalues of $ C_{mm}^T C_{mm} $ as $c_{m \alpha }  ~~~ (\alpha =1,\cdots M_m)$.
$ C_{mm}^T C_{mm} $ is a positive semidefinite matrix because $C_{mm}$ is a real matrix,
so $c_{m\alpha } \geq 0$.
Then we obtain $\delta \lambda_{m\gamma }^1$ from (\ref{eq:4-23}) and (\ref{eq:4-24}) .
\begin{equation}
\delta \lambda_{m\gamma }^1 = 
\begin{cases} 0 &  \\  %M_m -N_m  \\
\pm \dfrac{B_d}{r^{d-1}}  \sqrt{ c_{m\alpha } }  &  (\alpha =1,\cdots M_m) 
\end{cases} .
\label{eq:4-25}
\end{equation}
When $M_m < N_m$, we can obtain $\delta \lambda_{m\gamma }^1$ in the same way.
We define the eigenvalues of $ C_{mm} C_{mm}^T $ as $d_{m \alpha } (\geq 0) ~~~ (\alpha =1,\cdots N_m)$.
Then we obtain 
\begin{equation}
\delta \lambda_{m\gamma }^1 = 
\begin{cases} 0 &  \\  %M_m -N_m  \\
\pm \dfrac{B_d}{r^{d-1}}  \sqrt{ d_{m\alpha } }  &  (\alpha =1,\cdots N_m) 
\end{cases} .
\label{eq:4-25-2}
\end{equation}
Then $\sum_{m,\gamma } \delta \lambda_{m \gamma }^{1} \left. \dfrac{df}{d\lambda_m} \right|_{\lambda_m = \lambda_m^0} =0$, 
because we have $\sum_{\gamma}\delta \lambda_{m\gamma }^1=0$ from (\ref{eq:4-25}) and (\ref{eq:4-25-2}).
%cancel each other.

Next we consider $\delta \lambda_{m \gamma }^{2} $.
We skip the detailed calculation because it is also almost similar to the time-independent perturbation theory with degeneracy of quantum mechanics.
Then we can write $\delta \lambda_{m \gamma }^{2} $ as follows
\begin{equation}
\begin{split}
\delta \lambda_{m\gamma }^2  &= \sum_{n (\neq m) ,i, \beta } \dfrac{1}{\lambda_m^0 - \lambda_n^0} (f_{n i \beta }^{0T} A^0 \delta \Lambda_1 \xi_{m \gamma }^0 )
(\xi_{m \gamma }^{0T} A^0 \delta \Lambda_1 f_{n i \beta }^{0} )  \\
&= \sum_{n (\neq m) } \dfrac{1}{\lambda_m^0 - \lambda_n^0} \xi_{m \gamma }^{0T} A^0 \delta \Lambda_1 \hat{\phi }_n \delta \Lambda_1 \xi_{m \gamma }^0
\label{eq:4-26}
\end{split}
\end{equation}
where
\begin{equation}
 \hat{\phi }_n \equiv \sum_{ i, \beta }  f_{n i \beta }^{0} f_{n i \beta }^{0T} A^0 .
\label{eq:4-27}
\end{equation}
$ \hat{\phi }_n$ is a projection operator on the eigenspace of $\lambda_n^0$.
To obtain $\delta \lambda_{m \gamma }^{2} $ we must obtain $\xi_{m \gamma }^0$ by solving the eigenvalue problem,
 but it is not necessary for our purpose because we want to know only
$\sum_{m,\gamma } \delta \lambda_{m \gamma }^{2} \left. \dfrac{df}{d\lambda_m} \right|_{\lambda_m = \lambda_m^0} $.
From (\ref{eq:4-26}) we obtain
\begin{equation}
\begin{split} 
\sum_{m,\gamma } \delta \lambda_{m \gamma }^{2} \left. \dfrac{df}{d\lambda_m} \right|_{\lambda_m = \lambda_m^0} &=
\sum_{m,\gamma } \sum_{n (\neq m) } \dfrac{1}{\lambda_m^0 - \lambda_n^0} \xi_{m \gamma }^{0T} A^0 \delta \Lambda_1 \hat{\phi }_n \delta \Lambda_1 \xi_{m \gamma }^0
\left. \dfrac{df}{d\lambda_m} \right|_{\lambda_m = \lambda_m^0} \\
&=\sum_{m,n (m\neq n)} \dfrac{1}{\lambda_m^0 - \lambda_n^0} Tr ( \hat{\phi }_m \delta \Lambda_1 \hat{\phi }_n \delta \Lambda_1 )
\left. \dfrac{df}{d\lambda_m} \right|_{\lambda_m = \lambda_m^0} \\
&=\sum_{m,n (m>n)} \dfrac{1}{\lambda_m^0 - \lambda_n^0} Tr ( \hat{\phi }_m \delta \Lambda_1 \hat{\phi }_n \delta \Lambda_1 )
 \left( \left. \dfrac{df}{d\lambda_m} \right|_{\lambda_m = \lambda_m^0} -\left. \dfrac{df}{d\lambda_n} \right|_{\lambda_n = \lambda_n^0}  \right) .
\label{eq:4-28}
\end{split}
\end{equation}
In the second line we have used
\begin{equation}
\sum_{ \gamma  }  \xi_{m \gamma  }^{0} \xi_{m \gamma }^{0T} A^0 = \hat{\phi }_m ,
\label{eq:4-30}
\end{equation}
and in the third line we have used cyclic property of trace.
Next we examine the sign of (\ref{eq:4-28}).
Its trace term is positive because
\begin{equation}
\begin{split}
Tr ( \hat{\phi }_m \delta \Lambda_1 \hat{\phi }_n \delta \Lambda_1 )
&= \sum_{i ,\alpha ,j,\beta } (f_{n i \alpha  }^{0T} A^0 \delta \Lambda_1 f_{m j \beta  }^0 )
(f_{m j\beta  }^{0T} A^0 \delta \Lambda_1 f_{n i \alpha  }^{0} )  \\
&= \sum_{i ,\alpha ,j,\beta } V_{n\alpha m\beta }^{ij} V_{m\beta  n\alpha  }^{ji} 
= \sum_{i ,\alpha ,j,\beta } (V_{n\alpha m\beta }^{ij} )^2
= 2 \left(  \dfrac{B_d}{r^{d-1}}  \right) ^2  \sum_{\alpha ,\beta } (C_{m\alpha n\beta } )^2 \geq 0 .
\label{eq:4-31}
\end{split}
\end{equation} 
And from (\ref{eq:3-7}) we obtain
\begin{align}
\dfrac{df}{d\lambda } &=\dfrac{1}{2\sqrt{1+\lambda }} \ln \left[ \sqrt{1+\dfrac{1}{\lambda } } + \dfrac{1}{\sqrt{\lambda }} \right] >0 ~~~\textrm{for} ~~\lambda >0  , \label{eq:4-32} \\
\dfrac{d^2f}{d\lambda ^2 } &= - \dfrac{1}{4 \sqrt{1+\lambda }} \left[ \dfrac{1}{1+\lambda } \ln \left[ \sqrt{1+\dfrac{1}{\lambda } } + \dfrac{1}{\sqrt{\lambda }} \right]
+\dfrac{1}{\lambda \sqrt{1+\lambda }} \right]  < 0 ~~~\textrm{for} ~~\lambda >0  .
\label{eq:4-33}
\end{align} 
From (\ref{eq:4-31}), (\ref{eq:4-33}) and $\lambda_m^0 >\lambda_n^0 ~~(m>n) $ ,
(\ref{eq:4-28}) is negative.
And from  (\ref{eq:4-33}) we obtain
\begin{equation}
\sum_{m,\gamma } (\delta \lambda_{m \gamma }^{1} )^2 \left. \dfrac{d^2f}{d\lambda_m^2} \right|_{\lambda_m = \lambda_m^0} \leq 0  .
\end{equation}

Finally, from (\ref{eq:4-13}), (\ref{eq:4-25}), (\ref{eq:4-25-2}), (\ref{eq:4-28}) and (\ref{eq:4-31}) 
we obtain
\begin{equation}
\begin{split}
& S_{AB}(r,R_1,R_2) - S_A(R_1) - S_B(R_2) =  \sum_{m ,\gamma } 
\left[  \delta \lambda_{m\gamma }^2 \left. \dfrac{df}{d\lambda_m} \right|_{\lambda_m = \lambda_m^0} 
+\dfrac{1}{2} (\delta \lambda_m ^1 )^2 \left. \dfrac{d^2f}{d\lambda_m^2} \right|_{\lambda_m = \lambda_m^0} \right]  \\
&= \left(  \dfrac{B_d}{r^{d-1}}  \right) ^2 \Bigl[ \sum_{m,n (m>n)} \dfrac{2}{\lambda_m^0 - \lambda_n^0}   \sum_{\alpha ,\beta } (C_{m\alpha n\beta } )^2
 \left( \left. \dfrac{df}{d\lambda_m} \right|_{\lambda_m = \lambda_m^0} -\left. \dfrac{df}{d\lambda_n} \right|_{\lambda_n = \lambda_n^0}  \right)  \\
 & + \sum_{m',\alpha } c_{m'\alpha } \left. \dfrac{d^2f}{d\lambda_m^2} \right|_{\lambda_m = \lambda_{m'} ^0}
  + \sum_{m'',\alpha } d_{m''\alpha } \left. \dfrac{d^2f}{d\lambda_m^2} \right|_{\lambda_m = \lambda_{m''} ^0} \Bigr]  \equiv \dfrac{1}{r^{2d-2}} G(R_1,R_2) \leq 0
\label{eq:4-34}
\end{split}
\end{equation}
where $\sum_{m'}$ %sums
denotes the summation taken over the  common eigenvalues of $\Lambda^{01}$ and $\Lambda^{02}$, whose degeneracy is $M_m \geq  N_m$, 
and $\sum_{m''}$ %sums 
denotes the summation taken over the common eigenvalues of $\Lambda^{01}$ and $\Lambda^{02}$, whose degeneracy is $M_m < N_m$.
%From (\ref{eq:4-34}) $S_C(r,R_1,R_2) = S_{AB} (r,R_1,R_2)$ reaches its maximum value $S_A(R_1) +S_B(R_2)$ when $r$

We have obtained the $r$ dependence of $S_C(r,R_1,R_2) = S_{AB} (r,R_1,R_2)$ in (\ref{eq:4-34}), 
then we next consider $G(R_1,R_2)$.
To calculate $G(R_1,R_2)$ we need to know $\lambda_m^0$ and $f_{mi\alpha }^0$ which we do not examine in this paper.
But from $C_{m\alpha n\beta } (R_1=0,R_2) = C_{m\alpha n\beta } (R_1,R_2=0) =0$ we obtain a trivial property of $G(R_1,R_2)$,
\begin{equation}
G(R_1=0,R_2) =G(R_1,R_2=0) =0.
\label{eq:4-35}
\end{equation}    
And $G(R_1,R_2)$ depends on the cutoff length $l$ because $\lambda_m^0 ~,~ f_{mi\alpha }^0$ and $C_{m\alpha n\beta }$ depend on $l$.
($\lambda_m^0$ are dimensionless, so they depend on $R_1/l$ or $R_2/l$.  
And in (\ref{eq:4-22}) $\int_{A} d^d z_a W(x_a, z_a)$ and $\int_{B} d^d z_b W(y_b, z_b)$ depend on $l$ because $W(x,y)$ depend on $l$ for $x\approx y$,
  so $C_{m\alpha n\beta }$ depends on $l$. )
Probably $G(R_1,R_2)$ diverges when $l\rightarrow 0$,  as $S_A(R_1)$ and $S_B(R_2)$ have $1/l^{d-1}$ divergence \cite{Bombelli:1986rw,Srednicki:1993im}.
And %probably 
$G(R_1,R_2)$ most likely diverges more weakly than  $S_A(R_1)$ and $S_B(R_2)$. 
%Then we cannot fix the form of $G(R_1,R_2)$ using only dimensional analysis even if $R_1=R_2\equiv R$.
Then, by dimensional analysis, when $R_1=R_2\equiv R$ we can assume 
\begin{equation}
G(R_1=R,R_2=R) = g R^{2d-2} \left( \dfrac{R}{l} \right) ^m \left( \ln \left( \dfrac{R}{l} \right) \right) ^n  ~~~~ d-1 \geq m\geq 0 , n\geq 0, g<0
\label{eq:4-36}
\end{equation}    
%where $m$ and $n$ do not become zero simultaneously, and $g$ is a dimensionless constant.
where $g$ is a dimensionless constant.

%Next we consider the shapes of $A$ and $B$.
%In the above analysis we do not use the shapes of $A$ and $B$, so 
%all analysis in this section hold for $A$ and $B$ which have arbitrary shapes.
%In this case $R_1$ and $R_2$ are the characteristic sizes of $A$ and $B$.

Finally we consider the condition under which the approximations are good.
When $r\gg R_1,R_2$, ~   $\delta \Lambda \approx \delta \Lambda_1$ is a good approximation.
When 
$|\dfrac{ B_d}{ r^{d-1} } C_{ m \alpha n \beta }|  
\ll  |\lambda_m^0 -\lambda_n^0| $,
the perturbation theory is a good approximation.  
The latter condition might have $l$ dependence, so we might need 
the condition $R/r \ll (l/R )^a  $, where $a\geq 0$.

% figures should be put into the text as floats.
% Use the graphics or graphicx packages (distributed with LaTeX2e)
% and the \includegraphics macro defined in those packages.
% See the LaTeX Graphics Companion by Michel Goosens, Sebastian Rahtz,
% entire page. There is no need to do explicit centering.

% Surround figure environment with turnpage environment for landscape
% figure
% \begin{figure}
% \end{figure}

% \begin{figure}
% \end{figure}

%\section{two eternal black holes}

\begin{figure}
 \includegraphics[width=7cm,angle=270,clip]{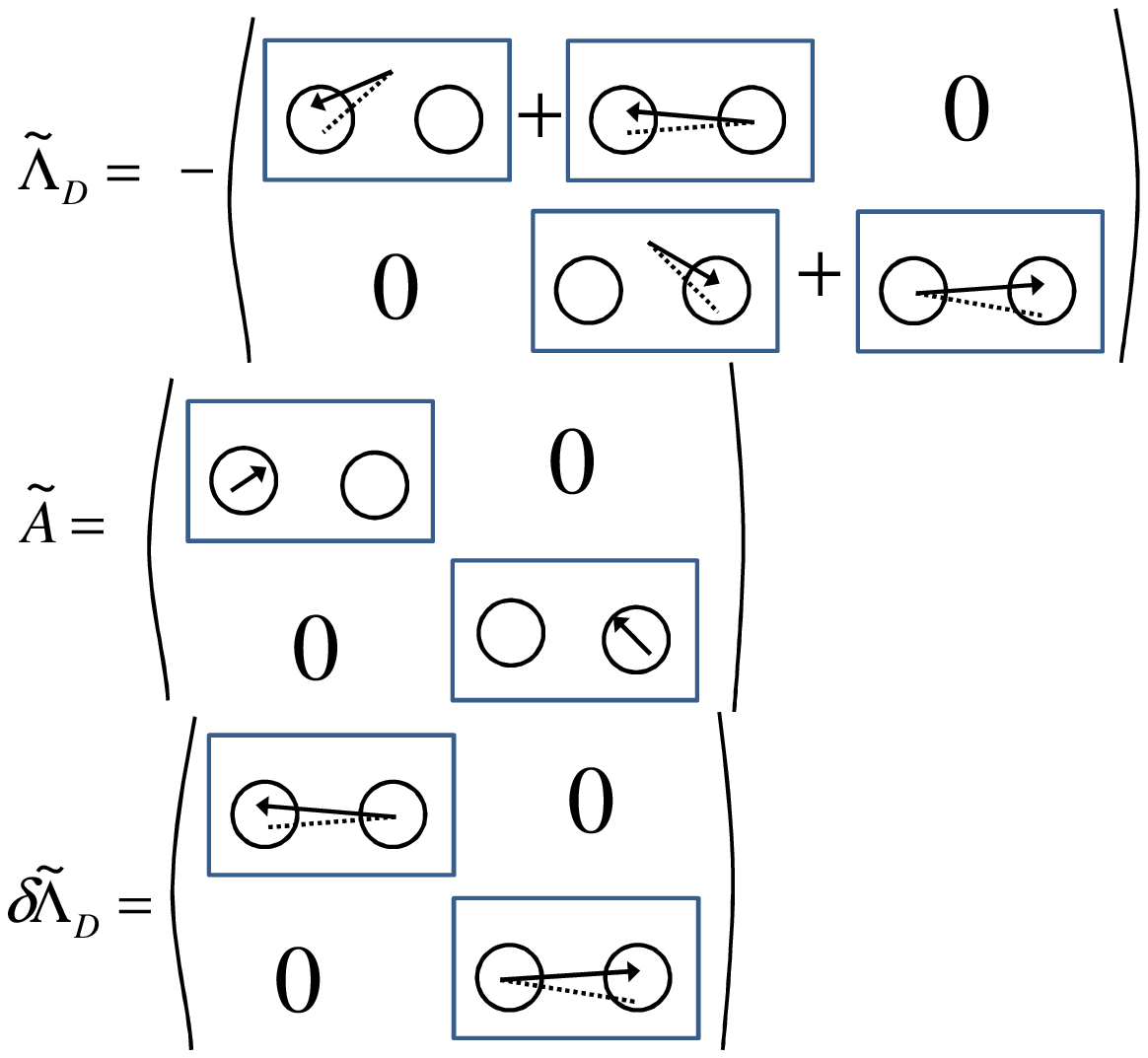}%
 \caption{The diagrammatic representations of $\tilde{\Lambda}_D$, $\tilde{A}$ and $ \delta \tilde{\Lambda}_D$.}
 \label{lambdatilde}
 \end{figure}

\section{entanglement entropy of two black holes in a $d+1$ dimensional massless free scalar field} \label{BH}
In this section we consider the  entanglement entropy of the massless free scalar field on the outside region $C$ of two black holes $A$ and $B$ whose radii are $R_1$ and $R_2$.
The action of the massless free scalar field is given by
\begin{equation}
S= - \dfrac{1}{2}\int d^d x \sqrt{-g} g^{\mu \nu } \nabla_\mu \phi  \nabla_\nu  \phi . 
\label{eq:new5-1}
\end{equation}
First we specify the vacuum state of the scalar field.
The vacuum state is specified by specifying the time coordinate $t$.
We use the coordinate system which have following properties:
this coordinate system covers the inside and the outside regions of two black holes and 
does not have the coordinate singularity on the horizons 
and becomes the orthogonal coordinate system of Minkowski spacetime in the region far from the two black holes.
To construct this coordinate system,  we use the coordinates which is similar to the Kruskal coordinates in the inside regions and the neighborhood  of black holes,
 and similar to the Schwarzschild coordinates in the other region.
In this coordinate system $g^{tt}$ is positive everywhere, then 
from (\ref{eq:new5-1}) $G_{MN}$ and $V_{MN}$ in (\ref{eq:3-1}) are positive definite. 
So we can use the formalism in the Section \ref{review}.

We can use the method of the last section with some modifications.
%We consider $S_{AB}(r,R_1,R_2)$ as the last section.
In the black hole case $W(x,y)$ and $W^{-1}(x,y)$ depend on $r$, 
so we write them as  $W(x,y;r)$ and $W^{-1}(x,y;r)$.
%As well as 
Exactly in the same way as  in Minkowski spacetime,   Eqs. (\ref{eq:new4-4})-(\ref{eq:new4-7}) hold because (\ref{eq:new4-2}) holds.
On the other hand $\Lambda_0 (=\lim_{r\rightarrow \infty} \Lambda)$ changes because  $W(x,y;r)$ and $W^{-1}(x,y;r)$ depend on $r$.
We define $W_A(x,y)$ and $W_A^{-1}(x,y)$($W_B(x,y)$ and $W_B^{-1}(x,y)$) as $W(x,y)$ and $W^{-1}(x,y)$ in the case that the only one black hole $A$($B$) exists.
Then we have 
\begin{equation}
 \Lambda^0 = - \begin{pmatrix}  \int_{A^c} d^{d} z W_A^{-1} (x_a,z) W_{A} (z,y_a) & 0  \\
                            0  &  \int_{B^c} d^{d} z  W_{B}^{-1} (x_b,z)  W_{B}(z,y_b)   \end{pmatrix} .
\label{eq:new5-2}
\end{equation}
It is difficult to evaluate the $r$ dependence of $\delta \Lambda_D =\Lambda_D -\Lambda^0$ because   
it is difficult to evaluate $W(x,y;r) - W_{A(B)}(x,y)$ and $W^{-1}(x,y;r) - W^{-1}_{A(B)}(x,y)$.
So, in the black hole case we \textit{do not} consider $\Lambda_0$ as the non-perturbative part.
Instead we define $\tilde{\Lambda}_D(r)$ and $\tilde{A} (r)$ as (see Fig \ref{lambdatilde})
\begin{align}
&\tilde{\Lambda}_D (r) \equiv
\begin{pmatrix}  \Lambda_A (x_a,y_a;r) &  0   \\
                                 0  & \Lambda_B (x_b,y_b;r)  \end{pmatrix}  \notag \\
&\equiv \begin{pmatrix}  -\int_{A^c} d^d z W^{-1}(x_a,z;r)W(z,y_a;r) &  0   \\
                                 0  &  -\int_{B^c} d^d z W^{-1}(x_b,z;r)W(z,y_b;r)   \end{pmatrix} ,  \\
&\tilde{A} (r) \equiv
\begin{pmatrix}  W (x_a,y_a;r) &  0   \\
                                 0  & W (x_b,y_b;r)  \end{pmatrix}  ,
\label{eq:5-1-1}
\end{align}
%where $A^c \equiv \bar{A}+B+\bar{B}$ and $B^c \equiv \bar{A}+A+\bar{B}$. 
and we consider $\tilde{\Lambda}_D$ as the non-perturbative part.
Note that $\Lambda_A$ and $\Lambda_B$ are the matrices $\Lambda$ corresponding to $S_A(r,R_1,R_2)$ and $S_B(r,R_1,R_2)$.
So we will obtain $S_{AB}$ as the following form, $S_{AB} (r,R_1,R_2)=S_A(r,R_1,R_2)+ S_B(r,R_1,R_2) +\delta S_{AB} (r,R_1,R_2)$.
We calculate the leading term  of $\delta S_{AB} (r,R_1,R_2)$ with respect to $1/r$.

We define $\delta \tilde{\Lambda}_D \equiv \Lambda_D -\tilde{\Lambda}_D $ (see Fig \ref{lambdatilde}),
then we have $\Lambda = \tilde{\Lambda}_D + \delta \Lambda_1 + \delta \Lambda_2 +\delta \tilde{\Lambda}_D$.
To evaluate $\delta \Lambda_1$, $\delta \Lambda_2$ and $\delta \tilde{\Lambda}_D$, we evaluate $W(x_a,y_b;r)$ and $W^{-1}(x_a,y_b;r)$.
When $r\gg R_1,R_2$, by dimensional analysis we obtain $W(x_a,y_b;r) \approx \tfrac{A_d}{r^{d+1}} L_1 (R_1/r,R_2/r)$ and 
$W^{-1}(x_a,y_b;r) \approx \tfrac{B_d}{r^{d-1}} L_2 (R_1/r,R_2/r)$, where $L_1$ and $L_2$ are dimensionless functions of $R_1/r$ and $R_2/r$.
The space time becomes Minkowski space time when $R_1\rightarrow 0$ and $R_2\rightarrow 0$, so in this limit probably we have $L_1\rightarrow 1$ and $L_2\rightarrow 1$.
This limit is equivalent to $r\rightarrow \infty$, so we have $\lim_{r\rightarrow \infty} L_1=\lim_{r\rightarrow \infty} L_2=1$.
Then we obtain $\delta \Lambda_1= O(1/r^{d-1})$, $\delta \Lambda_2 = O(1/r^{d+1})$ and $\delta \tilde{\Lambda}_D = O(1/r^{2d})$ 
 as well  as the Minkowski spacetime case.
We can neglect $\delta \Lambda_2$ and $\delta \tilde{\Lambda}_D$ for the same reason 
as in the Minkowski spacetime case (see below Eq.(\ref{eq:new4-13})).
So we can approximate $\Lambda \approx  \tilde{\Lambda}_D + \delta \Lambda_1$.
Then we change the perturbative calculation in the last section as follow 
\begin{equation}
\Lambda^0  \rightarrow  \tilde{\Lambda}_D (r)
~~~~A^0  \rightarrow  \tilde{A} (r)
~~~~\lambda _m^0 \rightarrow \tilde{\lambda} _m^0 (r)
~~~~f_{mi\alpha }^0 \rightarrow {\tilde{f}_{mi\alpha } }^0 (r)
\label{eq:5-1}
\end{equation}
where $\tilde{\lambda} _m^0 (r) $ and $\tilde{f}_{mi\alpha }^0 (r) $ are the eigenvalues and the eigenvectors of $  \tilde{\Lambda}_D (r)$.
The perturbative calculation is the same as that in the last section.
In this case $\tilde{A} (r)$, $\tilde{\lambda} _m^0 (r) $ and ${\tilde{f}_{mi\alpha } }^0 (r)$ depend on $r$, but we can remove their $r$ dependence as follow. 
Because we want to calculate the leading term  of $S_{AB}(r,R_1,R_2) -S_A(r,R_1,R_2)-S_B(r,R_1,R_2)$ with respect to $1/r$,
we can approximate
\begin{equation}
 \begin{split}
 \tilde{A} \delta \Lambda_1 
& \approx 
 \dfrac{B_d L_2 (\tfrac{R_1}{r},\tfrac{R_2}{r})}{r^{d-1}}  
\begin{pmatrix}  0  & \int_{A} d^d z_a W(x_a, z_a;r)  \int_{B} d^d z_b W(z_b, y_b;r )     \\
                      \int_{B} d^d z_b W(x_b, z_b;r ) \int_{A} d^d z_a W(z_a, y_a;r)    &  0    \end{pmatrix}  \\
& \approx 
 \dfrac{B_d}{r^{d-1}}  
\begin{pmatrix}  0  & \int_{A} d^d z_a W_A(x_a, z_a)  \int_{B} d^d z_b W_B(z_b, y_b )     \\
                      \int_{B} d^d z_b W_B(x_b, z_b ) \int_{A} d^d z_a W_A(z_a, y_a)    &  0    \end{pmatrix} 
\end{split} .
\label{eq:new5-3}
\end{equation}
In the second line  we have approximated $L_2 (\tfrac{R_1}{r},\tfrac{R_2}{r}) \approx 1$, $W(x_a, z_a;r) \approx W_A(x_a, z_a)$
 and $W(z_b, y_b;r ) \approx W_B(z_b, y_b ) $.
And we can approximate $ \tilde{\lambda} _m^0 (r) \approx \tilde{\lambda}_m^0 (r=\infty) \equiv \lambda_m^0 $
 and  $ \tilde{f}_{mi\alpha }^0 (r) \approx  \tilde{f}_{mi\alpha }^0 (r=\infty) \equiv f_{mi\alpha }^0  $.
Note that $\lambda _m^0 $ and $f_{mi\alpha }^0 $ are the eigenvalues and the eigenvectors of $ \Lambda^0$, i.e.  ($\Lambda^0$ is in (\ref{eq:new5-2}))
\begin{equation}
f_{m1\alpha }^0 =
\begin{pmatrix} f_{m1\alpha }^0 (x_a)  \\
0   \end{pmatrix} ,
f_{n2\beta  }^0 =
\begin{pmatrix} 0 \\ f_{n2\beta  }^0 (x_b)  \end{pmatrix} ,
\Lambda^0 f_{m1\alpha }^0 = \lambda_m^0 f_{m1\alpha }^0, 
\Lambda^0 f_{m2\beta  }^0 = \lambda_m^0 f_{m2\beta  }^0, 
\label{eq:new5-4-2}
\end{equation} 
where $\alpha=1,\cdots, M_m $ and $\beta=1,\cdots,N_n $ are the labels of the degeneracy.

Finally we obtain
\begin{equation}
S_{AB}(r,R_1,R_2) = S_A(r,R_1,R_2) + S_B(r,R_1,R_2) + \dfrac{1}{r^{2d-2}} G(R_1,R_2) 
\label{eq:5-2}
\end{equation}
where $G(R_1,R_2)$ is the same function as that in (\ref{eq:4-34}).
Note that in this case from (\ref{eq:new5-3}) $C_{m\alpha n\beta }$ in $G(R_1,R_2)$ is
\begin{equation}
C_{m\alpha n\beta } =
\int_{A} d^dx_a \int_{A} d^d z_a W_A(x_a, z_a) f_{m1\alpha }^0 (x_a) \int_{B} d^dy_b \int_{B} d^d z_b W_B(y_b, z_b ) f_{n2\beta  }^0 (y_b) .
\label{eq:new5-4}
\end{equation}
As in the Minkowski spacetime case,
we obtain $G(R_1=0,R_2)=G(R_1,R_2=0)=0$ from $C_{m\alpha n\beta } (R_1=0,R_2) = C_{m\alpha n\beta } (R_1,R_2=0) =0$, 
and $G(R_1,R_2)$ probably diverges when $l\rightarrow 0$, where $l$ is the cutoff length.
The $1/l$ dependence of $G(R_1,R_2)$ is most likely the same as that in the Minkowski spacetime,
 then we obtain
 \begin{equation}
G(R_1=R,R_2=R) = g_{BH} R^{2d-2} \left( \dfrac{R}{l} \right) ^m \left( \ln \left( \dfrac{R}{l} \right) \right) ^n  ~~~~ d-1 \geq m\geq 0 , n\geq 0, g_{BH}<0
\label{eq:new5-5}
\end{equation}    
where $g_{BH}$ is a dimensionless constant, and $m$ and $n$ are the same numbers as those in the Minkowski spacetime.

\section{entanglement entropic force and the physical prediction} \label{prediction}
We \textit{assume} that we can consider the entanglement entropy of two black holes as thermodynamic entropy.
If this assumption is correct, the entropic force acts on two black holes.
We consider the force of the scalar field which acts on two black holes.
We consider two black holes which have same radius $R_1=R_2\equiv R$,
then we can consider the temperature $T$ to be the Hawking temperature.
We define the energy and the free energy of the field on the region $C$ as $E_C(r,R)$ and $F_C(r,R)$,
\begin{equation}
F_{C} (r,R) =E_C(r,R) - TS_{C}(r,R) =E_C(r,R) -T \left( 2S_A(r,R) +  \dfrac{1}{r^{2d-2}} G(R) \right) .
\label{eq:6-1}
\end{equation}
where $G(R) \equiv G(R_1=R,R_2=R)$ and we have used (\ref{eq:5-2}).      
We define the force of the field on the region $C$ 
which acts on one black hole in the direction of increasing $r$ as $X_C$.
We obtain $X_C$ by partially differentiating $F_C$ with $R$ fixed,
 \begin{equation}
X_{C} (r,R) =-\dfrac{\partial F_{C}}{ \partial r} = -\dfrac{ \partial E_C(r,R) }{\partial r} +T \left( 2\dfrac{\partial S_A(r,R)}{\partial r} -(2d-2)  \dfrac{1}{r^{2d-1}} G(R) \right) .
\label{eq:6-2}
\end{equation}
In (\ref{eq:6-2}) %the first term is the ordinary back reaction and 
the second term is the entropic force.

\begin{figure}
 \includegraphics[width=8cm,angle=270,clip]{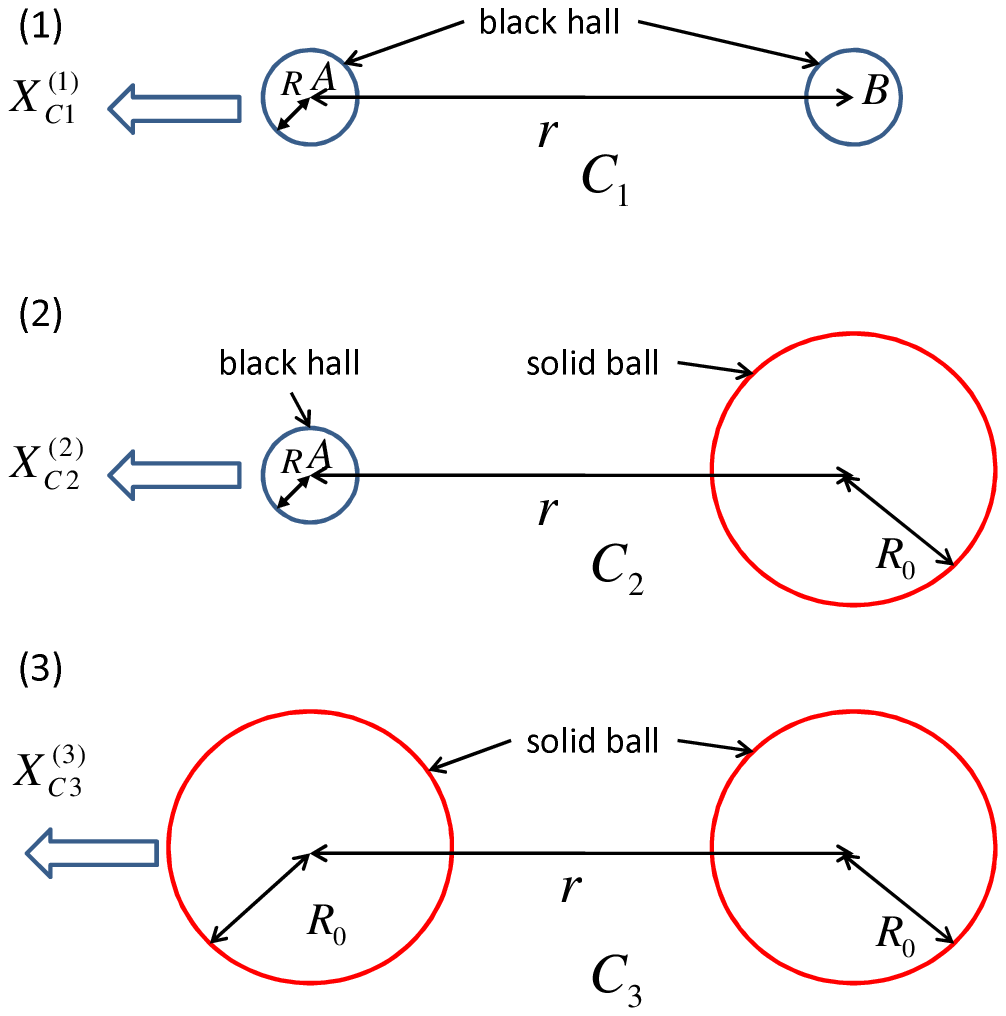}%
 \caption{Three situations to see the effect of the entropic force. 
(1) There are two black holes.  (2)  There are one black hole and one solid ball. (3) There are two solid balls. 
 We define the force of the field 
which acts on one black hole or on one ball in the direction of increasing $r$ as $X^{(1)}_C$, $X^{(2)}_{C_2}$ and $X^{(3)}_{C_3}$. }
 \label{situations}
 \end{figure}

We cannot see the effect of the entropic force only from (\ref{eq:6-2}) because we do not know $S_{A}(r,R)$.
 To see the effect of the entropic force we  consider three situations . (See Fig \ref{situations})
 (1) There are two black holes which have the same radius $R$ and the distance between them is $r$.(This is the situation we have considered.)
(2)  There are one black hole whose radius is $R$ and one solid ball whose radius is $R_0 \approx R (R_0 >R)$, and the distance between them is $r$. 
This ball has mass $M$ which is the same as that of a black hole whose radius is $R$.
And the scalar field does not exist in this ball. 
The boundary condition on the scalar field on the surface of this ball is not so important in the later calculation  
that we do not specify the boundary condition in detail. 
We only require that the scalar field on the outside region of this ball is not so different from that in the situation (1).
%one of them (B) is surrounded by a light spherical wall which radius is $R_0\approx R (R_0>R)$.
%(2)  There are two black holes which have the same radius $R$ and the distance between them is $r$ and 
%one of them (B) is surrounded by a light spherical wall which radius is $R_0\approx R (R_0>R)$.
(3) There are two solid balls which have the same radius $R_0$ and the distance between them is $r$. 
These balls have the same properties as those in the situation (2).

 We define the force of the field 
which acts on one black hole or on one ball in the direction of increasing $r$ as $X^{(1)}_C$, $X^{(2)}_{C_2}$ and $X^{(3)}_{C_3}$.
We illustrate in  Fig \ref{situations} the directions of force and the names of the regions. %in Fig \ref{situations}.
%We represent the directions of force and the names of the regions in Fig \ref{situations}.

In the situation (2)  the state of the field is $\ket{0}^{(2)}_{A+C_2 }$, where $\ket{0}^{(2)}_{ A+C_2 }$ is the vacuum state on $A+C_2 $.
%the tensor product, $\ket{0 }^{(2)}_{\mathbb{R}^d } =\ket{0}^{(2)}_{A+C_2 } \otimes  \ket{0 }^{(2)}_{B+B' }$, %the tensor product of the states of the fields on two regions
%where $\ket{0}^{(2)}_{ A+C_2 }$ and $\ket{0}^{(2)}_{B+B'}$ are the vacuum state on $A+C_2 $ and $B+B' $.
Because $\ket{0}^{(2)}_{ A+C_2 }$ is a pure state, then $S^{(2)}_{C_2} =S^{(2)}_{A}$. 
We define $\Lambda_A^{(1)}$($\Lambda_A^{(2)}$) as $\Lambda$ corresponding to $S^{(1)}_{A}$($S^{(2)}_{A}$).
Because the scalar field does not exist in the ball, then we obtain 
\begin{equation}
\Lambda_A^{(2)} -\Lambda_A^{(1)} \approx \int_B d^dz_b W^{-1}(x_a,z_b) W(z_b,y_a) =O\left( \dfrac{1}{r^{2d}} \right) .
\label{eq:new6-1}
\end{equation}
Then we can approximate $S^{(2)}_{A} =S^{(1)}_{A} (r,R) + O(\tfrac{1}{r^{2d}}) \approx S^{(1)}_{A} (r,R) $.
Then we obtain
\begin{equation}
X^{(2)}_{C_2} (r,R) =-\dfrac{\partial F^{(2)}_{C_2}}{ \partial r} = -\dfrac{ \partial E^{(2)}_{C_2} (r,R) }{\partial r} +T  \dfrac{\partial S^{(2)}_{A} (r,R)}{\partial r} 
\approx   -\dfrac{ \partial E^{(2)}_{C_2} (r,R) }{\partial r} +T  \dfrac{\partial S^{(1)}_{A} (r,R)}{\partial r} .
\label{eq:6-3}
\end{equation}

In the situation (3) the state of the field on the region $C_3$ is a pure state, so $S^{(3)}_{C_3} =0$.
Then we obtain
\begin{equation}
X^{(3)}_{C_3} (r,R) =-\dfrac{\partial F^{(3)}_{C_3}}{ \partial r} = -\dfrac{ \partial E^{(3)}_{C_3} (r,R) }{\partial r}  .
\label{eq:6-4}
\end{equation}

%Next we use the condition $R_0\approx R$.
%When $R_0\approx R$ we can approximate $ E^{(1)}_{C_1} \approx   E^{(2)}_{C_2}  \approx E^{(3)}_{C_3} $.
From (\ref{eq:6-2}) (\ref{eq:6-3}) and (\ref{eq:6-4}) we obtain
\begin{equation}
X^{(1)}_{C_1} - 2 X^{(2)}_{C_2} + X^{(3)}_{C_3}  \approx -\dfrac{\partial }{\partial r} [E^{(1)}_{C_1} -2E^{(2)}_{C_2} + E^{(3)}_{C_3}] -(2d-2) T \dfrac{1}{r^{2d-1}} G(R) .
\label{eq:6-5}
\end{equation}
$E^{(1)}_{C_1} -2E^{(2)}_{C_2} + E^{(3)}_{C_3}$ is Casimir energy.

We have not considered the force of gravity. 
But we can include them in (\ref{eq:6-5}) easily.
We define total force acting on one black hole or on one ball 
in the direction of increasing $r$ as   $\mathcal{F}^{(1)}_A  $, $\mathcal{F}^{(2)}_A  $ and $\mathcal{F}^{(3)}_{ball} $. %, where $G_N$ is the Newton constant.-\tfrac{G_N M^2}{r^2}-\tfrac{G_N M^2}{r^2 }-\tfrac{G_N M^2}{r^2}
Then we obtain
\begin{equation}
\mathcal{F}^{(1)}_A -2 \mathcal{F}^{(2)}_A + \mathcal{F}^{(3)}_{ball} =X^{(1)}_{C_1} - 2 X^{(2)}_{C_2} - X^{(3)}_{C_3}  
\approx -\dfrac{\partial }{\partial r} [E^{(1)}_{C_1} -2E^{(2)}_{C_2} + E^{(3)}_{C_3}] -(2d-2) T \dfrac{1}{r^{2d-1}} G(R) .
\label{eq:6-6}
\end{equation}
The force of gravity is canceled in (\ref{eq:6-6}).
The first and the second terms in the right hand side are the Casimir force and the effect of entropic force, respectively.
%The left hand side of (\ref{eq:6-6}) can be measured experimentally, so  (\ref{eq:6-6}) is the physical prediction.

Finally we consider the case $d=3$.
In this case the Hawking temperature is $T=\tfrac{1}{8\pi G_N M} =\tfrac{1}{4\pi R}$.
From (\ref{eq:new5-5}) and (\ref{eq:6-6}) we obtain
\begin{equation}
\begin{split}
& \mathcal{F}^{(1)}_A -2 \mathcal{F}^{(2)}_A + \mathcal{F}^{(3)}_{ball}  \approx -\dfrac{\partial }{\partial r} [E^{(1)}_{C_1} -2E^{(2)}_{C_2} + E^{(3)}_{C_3}] 
  -\dfrac{g_{BH}}{\pi}  \dfrac{R^3}{r^{5}} 
 \left( \dfrac{R}{l} \right) ^m \left( \ln \left( \dfrac{R}{l} \right) \right) ^n \\
& 2 \geq m\geq 0 , n\geq 0, g_{BH}<0 .
\label{eq:6-7}
\end{split}
\end{equation}
We roughly estimate the Casimir force by analogy with that of electromagnetic field between two dielectric spheres with center-to-center distance $r$ in Minkowski spacetime.
The Casimir force between the two sphere was calculated in \cite{Emig:2007cf}, and it is $O(1/r^8)$.
So, in our case we can probably neglect $-\tfrac{\partial }{\partial r} [E^{(1)}_{C_1} -2E^{(2)}_{C_2} + E^{(3)}_{C_3}] $  in (\ref{eq:6-7}). 
The left hand side of (\ref{eq:6-7}) can be measured experimentally, so  (\ref{eq:6-7}) is the physical prediction.
From (\ref{eq:6-7}) the effect of the entropic force becomes significant when $R$ is large. 
We can probably use heavy stars as the balls in the situation (2) and (3).
So  we can possibly confirm the effect of the entropic force by the cosmic observation (e.g. binary black holes and binary neutron stars). 
   
We estimate the magnitude of the effect of the entropic force.
We set the cutoff length $l$ to the Planck length $l_P =(G_N \hbar /c^3)^{1/2}$, 
then the ratio of the effect of the entropic force to the force of gravity is
\begin{equation}
\dfrac{\mathcal{F}_{eef} }{ \mathcal{F}_{g}} =\dfrac{4g_{BH}}{\pi} \dfrac{(l_P)^2 R}{r^3} \left( \dfrac{R}{l_P} \right) ^m \left( \ln \left( \dfrac{R}{l_P} \right) \right) ^n
\label{eq:6-8}
\end{equation}
where
\begin{equation}
\mathcal{F}_{eef} \equiv -\dfrac{\hbar c g_{BH}}{\pi}  \dfrac{R^3}{r^{5}}  \left( \dfrac{R}{l_P} \right) ^m \left( \ln \left( \dfrac{R}{l_P} \right) \right) ^n
~~~~ \mathcal{F}_{g} \equiv - \dfrac{G_N M^2}{r^2} = -\dfrac{c^4 R^2}{4G_N r^2} .
\label{eq:6-9}
\end{equation}
If $m=2$, $l_P$ is canceled in (\ref{eq:6-8}). 
In this case the effect of the entropic force is comparable to the force of gravity, 
so we can possibly observe the effect of the entropic force.

%We can probably use plasma and heavy stars as the wall and the balls in the situation (2) and (3).
%So  we can possibly confirm the effect of the entropic force by the cosmic observation. 
   
%\cite{Emig:2007cf}

\section{conclusion and discussion}
In Section \ref{general} we showed that the entanglement entropy ($S_C=S_{AB}$) of two disjoint regions in
translational invariant vacuum in general QFT reaches its maximum value when $r\rightarrow \infty$.
And we obtained the inequality (\ref{eq:2-2}).
In Section \ref{flat} we developed the method to obtain the $r$ dependence of $S_C$ 
and obtained the $r$ dependence of $S_C$ (\ref{eq:4-34}) in the free massless scalar field in $(d+1)$ dimensional Minkowski spacetime.
We can use this method in curved space time and for scalar field theory whose Lagrangian is quadratic.
To know only the $r$ dependence we need only the $\| x-y\| $ dependence of $W(x,y)$ and $W^{-1}(x,y)$ when $\|x-y\|$ is large.
To know the $R_1$ and $R_2$ dependence we must solve the zeroth order eigenvalue equation and obtain $ \lambda^{0}_m$  and $f^0_{ m i\alpha }$.     
It is difficult to solve the zeroth order eigenvalue equation analytically, so we will need to perform numerical calculation.
But we assumed the $R_1$ and $R_2$ dependence (\ref{eq:4-36}) by using dimensional analysis and the cutoff dependence of $S_A$ and $S_B$.
  
In Section \ref{BH} 
we showed that $S_C$ can be expected to be the form (\ref{eq:5-2}) in the black hole case. 
In this case the only assumption we made is the $r$ dependence of $W(x_a,y_b)$ and $W^{-1}(x_a,y_b)$.
 We did not explicitly calculate $W(x_a,y_b)$ and $W^{-1}(x_a,y_b)$, but  
assumed the $r$ dependence of $W(x_a,y_b)$ and $W^{-1}(x_a,y_b)$ by dimensional analysis.

In Section \ref{prediction} we assumed that we can consider the entanglement entropy of two black holes as thermodynamic entropy, 
and investigated its entropic force.
We considered three situations (1), (2) and (3) and obtain the relationship (\ref{eq:6-7}) between the force acting on one black hole or on one ball 
and the sum of the Casimir force and the effect of the entanglement entropic force.
Because we can probably neglect the Casimir force, 
we can confirm (\ref{eq:6-7}) experimentally in principle. 
And we can possibly confirm the effect of the entropic force by the cosmic observation
because it is significant for large black holes.

Next we discuss the entanglement entropic force in different systems.
In the black hole case, black holes act as \textit{"walls" which hide inside regions but hold the entanglement 
between inside and outside regions}.
So if there are walls of this type, the entanglement entropic force will exist between regions surrounded by these walls.  
Then we will be able to confirm the entanglement entropic force by experiments in a laboratory 
if we make this wall.
And if entanglement entropy depends on some external parameter, 
 entanglement entropic force probably appears also in quantum mechanical (i.e. not quantum field theoretical) systems. 

Finally we mention our assumption that we can consider the entanglement entropy of two black holes as thermodynamic entropy.
Entanglement entropy has  property which is different from that of thermodynamic entropy.
For example entanglement entropy is not a extensive variable in general.
So we must reconsider statistical mechanics from a fundamental level to judge 
whether our assumption is correct or not.
We can also use (\ref{eq:6-7}) to judge the correctness of our assumption by experiments.

\begin{acknowledgments}
I am grateful to Takahiro Kubota and Satoshi Yamaguchi  for a careful reading of
this manuscript and useful comments and discussions.
I also would like to thank Yutaka Hosotani and Kin-ya Oda for useful discussions.
This work was supported in part by JSPS Research Fellowship for Young
Scientists.
\end{acknowledgments}


\appendix
\section{The calculation of $W$ and $W^{-1}$}
In this appendix we calculate $W(x,y)$ and $W^{-1}(x,y)$ ( (\ref{eq:3-10}) and (\ref{eq:3-11}) ) explicitly.
We regularize them by including convergence factor $e^{-l|k|}$ in them, where $l$ is the cutoff length.
We define $W_\alpha $ as 
\begin{equation}
W_\alpha (x,y) = \int \dfrac{d^d k}{ (2\pi )^d } (k^2  )^{(1-\alpha ) /2}  e^{ik \cdot (x-y)} e^{-l|k|} .  \label{eq:A-1} 
\end{equation}
Then we have $W_0=W$ and $W_2=W^{-1}$.
First we consider the case $d\geq 3$.

(1) $d\geq 3$

We perform the integrals of angular coordinates which do not enter the inner product,
\begin{equation}
W_\alpha (x,y) = \dfrac{1}{(2\pi )^d} \prod_{m-2}^{d-2} \left( \sqrt{\pi} \dfrac{\Gamma \left( \dfrac{d-m}{2} \right) }{ \Gamma \left( \dfrac{d-m+1}{2} \right) } \right)
\int_{0}^{\infty} dk \int_{-1}^{1} dt [1-t^2]^{\tfrac{d-3}{2}}  e^{ik rt} e^{-lk} k^{d-\alpha }  
\label{eq:A-2} 
\end{equation}
where $r\equiv \| x-y \| $ and we change the variable as $t=\cos \theta $.
Next we perform the $k$ integral
\begin{equation}
\begin{split}
& \int_{0}^{\infty} dk \int_{-1}^{1} dt [1-t^2]^{\tfrac{d-3}{2}}  e^{ik rt} e^{-lk} k^{d-\alpha } 
=  \int_{-1}^{1} dt [1-t^2]^{\tfrac{d-3}{2}}  \left( \dfrac{1}{it} \dfrac{d}{dr} \right) ^{d-\alpha } \int_{0}^{\infty} dk e^{ik rt} e^{-lk} \\
&= (-i)^{d-\alpha -1 }  (d-\alpha ) ! \dfrac{1}{r^{d-\alpha +1 }} \int_{-1}^{1} dt [1-t^2]^{\tfrac{d-3}{2}} \dfrac{1}{ (t+iz)^{d-\alpha +1} } 
\label{eq:A-3}
\end{split} 
\end{equation}
where $z \equiv l/r $. 
We define
\begin{equation}
g(t) \equiv  [1-t^2]^{\tfrac{d-3}{2}} \dfrac{1}{ (t+iz)^{d-\alpha +1} } .
\label{eq:A-4}
\end{equation}
We want to show $W_\alpha \neq 0$ when $z\rightarrow 0$.
%We first consider the case $d=2m+2 ~~(m\geq 1)$.

\begin{figure}
 \includegraphics[width=8cm,angle=270,clip]{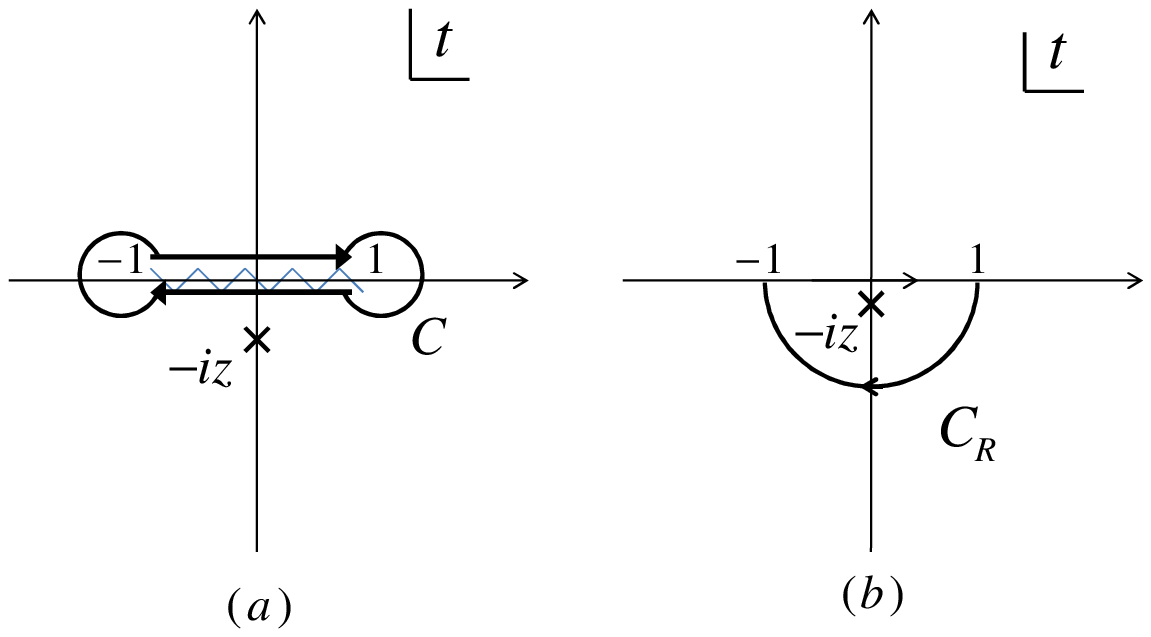}%
 \caption{The contours of the integrals. (a) $d=2m+2 ~~(m\geq 1)$. (b) $d=2m+1 ~~(m\geq 1)$. }
 \label{integral}
 \end{figure}

(i) $d=2m+2 ~~(m\geq 1)$

In this case $g(t)$ has a branch cut on the real axis from $-1$ to $1$.
We perform the integration along the contour shown in Fig \ref{integral} (a), and obtain
%We perform the contour integral, and we obtain (the contour is shown in Fig \ref{integral} (a))
\begin{equation}
\int_{-1}^{1} dt g(t) = \pi i \mathrm{Res}_{t=-iz}  g(t) =\pi i \dfrac{1}{ (d-\alpha ) !} \left. \left( \dfrac{d^{d-\alpha }}{dt^{d-\alpha } }  [1-t^2]^{\tfrac{d-3}{2}} \right) \right| _{t=-iz} .
\label{eq:A-5}
\end{equation}
The derivative in (\ref{eq:A-5}) can be calculated by the derivative of a composite function,
\begin{equation}
 \begin{split}
  \left( \dfrac{d^{d-\alpha }}{dt^{d-\alpha } }  [1-t^2]^{\tfrac{d-3}{2}} \right) 
  =&  \sum_{r=0}^{\left[ \tfrac{1}{2} (d- \alpha) \right] } \dfrac{ (d-\alpha)! }{r! (d-\alpha -2r)!} (2t)^{d-\alpha-2r} 
    (-1)^{d-\alpha-r}   \\ 
   & (\dfrac{d-3}{2}) (\dfrac{d-3}{2} -1) \cdots (\dfrac{d-3}{2}- (d-\alpha-r-1))
    (1-t^2)^{\tfrac{d-3}{2}  -(d-\alpha -r)} 
\label{eq:A-6}
\end{split}
\end{equation}
where $\left[ \tfrac{1}{2} (d- \alpha) \right]$ is the Gauss' symbol which is the greatest integer that is less than or equal to $\tfrac{1}{2} (d- \alpha)$.

Then, when $z\rightarrow 0$, we obtain  %note that the branch cut so (-1) appear 
\begin{equation}
\begin{split}
\int_{-1}^{1} dt g(t) &= -\pi i  \dfrac{ 1 }{ (\tfrac{ d-\alpha }{2})! } 
    (-1)^{\tfrac{d-\alpha}{2}}  
    (\dfrac{d-3}{2}) (\dfrac{d-3}{2} -1) \cdots (\dfrac{d-3}{2}- (\dfrac{d-\alpha}{2} -1)) \\
   &= - \pi i  \dfrac{ 1 }{ (\tfrac{ d-\alpha }{2})! } 
    \left(-\dfrac{1}{2} \right)^{\tfrac{d-\alpha}{2}}  
    (d-3) (d-1) \cdots ( \alpha -1) .
   \label{eq:A-7}
 \end{split}  
\end{equation}
Then, from (\ref{eq:A-2}), (\ref{eq:A-3}) and (\ref{eq:A-7}), 
for $\alpha =2l \leq d (l \in \mathbb{Z})$  $W_\alpha $ is nonzero and $W_\alpha$ has the form of (\ref{eq:4-8})
when  $z\rightarrow 0$.
(When $\alpha=d$, we obtain $\int_{-1}^{1} dt g(t) =-\pi i$ from (\ref{eq:A-5}). 
(Note that $g(t)$ has the branch cut, then $[1+z^2]^{(d-3)/2} \rightarrow (-1)$ when $z\rightarrow 0$.))
%And we obtain

(ii) $d=2m+1 ~~(m\geq 1)$

We perform the integration along the contour shown in Fig \ref{integral} (b), and obtain
%We perform the contour integral, and we obtain (the contour is shown in Fig \ref{integral} (b))
\begin{equation}
\int_{-1}^{1} dt g(t) = -2 \pi i \mathrm{Res}_{t=-iz}  g(t)  - \int_{C_R} dt g(t) .
\label{eq:A-8}
\end{equation}
For $\alpha =2l  (l \in \mathbb{Z})$, $d-\alpha$ is odd, so we obtain $\lim_{z\rightarrow 0} \mathrm{Res}_{t=-iz}  g(t) =0$ from (\ref{eq:A-6}).
Then, for $\alpha =2l$ and $z\rightarrow 0$ we obtain
\begin{equation}
\begin{split}
\int_{-1}^{1} dt g(t) &=  - \int_{C_R} dt g(t) 
= i (-1)^{d-\alpha} \int_0^{\pi} d\theta e^{-i(d-\alpha )\theta  } [1-e^{2i\theta}]^{\tfrac{d-3}{2}}   \\
&= i (-1)^{d-\alpha}  (-2i)^{ m-1 }  \int_0^{\pi} d\theta e^{-i(2m+1-\alpha )\theta  } e^{i(m-1)\theta } (\sin \theta )^{m-1}  \\
&= (-1)^m 2^{m-1} i^m ( I _{sc} [m-1,m+2-\alpha ] -i I_{ss} [m-1,m+2-\alpha ] )   \\
&= \begin{cases} 2^{m-1} i^{m+1} I_{ss} [m-1,m+2-\alpha ] \neq 0  & \textrm{for odd} ~~ m  , \\
                       2^{m-1} i^{m} I_{sc} [m-1,m+2-\alpha ] \neq 0 & \textrm{for even} ~~ m   ,
     \end{cases}     
\label{eq:A-9}
\end{split}
\end{equation}
where 
\begin{equation}
 I _{sc} [m,n ] \equiv \int_0^{\pi} d\theta  (\sin \theta )^{m} \cos (n\theta ) ~, ~~~
  I _{ss} [m,n ] \equiv \int_0^{\pi} d\theta  (\sin \theta )^{m} \sin (n\theta)  .
\label{eq:A-10}
\end{equation}
Then, from (\ref{eq:A-2}) (\ref{eq:A-7}) and (\ref{eq:A-9}) ,  
for $\alpha =2l  (l \in \mathbb{Z})$  $W_\alpha $ is nonzero and $W_\alpha$ has the form of (\ref{eq:4-8})
when  $z\rightarrow 0$.

From (i) and (ii) we showed  (\ref{eq:4-8}) for $d\geq 3$.
Next we consider $d=2$.

(2) $d=2$

In this case we can perform the angular integral first,
\begin{equation}
\begin{split}
W_\alpha (x,y) &= \dfrac{1}{(2\pi )^2}  \int_0^{\infty} kdk \int_0^{2\pi} d\theta k^{1-\alpha} e^{ikr \cos \theta -lk}  \\
&= \dfrac{1}{ 2\pi }  \int_0^{\infty} dk k^{2-\alpha} e^{-lk} J_0 (kr)  
=  \dfrac{1}{ 2\pi r^{3-\alpha} } \int_0^{\infty} dx x^{2-\alpha} e^{-zx} J_0 (x) 
\label{eq:A-11}
\end{split} 
\end{equation}
where $J_0$ is the Bessel function of zeroth order.
We perform the integral for $\alpha=2$ and $\alpha=0$.

(i) $\alpha=2$

In this case we have
\begin{equation}
\int_0^{\infty} dx  e^{-zx} J_0 (x) = \dfrac{1}{\sqrt{z^2 +1}} .
\label{eq:A-12}
\end{equation}
Then, when  $z\rightarrow 0$ we obtain
 \begin{equation}
W_{\alpha =2} (x,y) = W^{-1} (x,y) = \dfrac{1}{2\pi r}  .
\label{eq:A-13}
\end{equation}

(i) $\alpha=0$

In this case we have
\begin{equation}
\int_0^{\infty} dx x^2 e^{-zx} J_0 (x) = \dfrac{ \Gamma(3) }{ z^3} F\left( \dfrac{3}{2} ,2,1;-\dfrac{1}{z^2} \right) \rightarrow -1  ~~(z\rightarrow 0) 
\label{eq:A-14}
\end{equation}
where $F$ is the Gaussian hypergeometric function.
Then, when  $z\rightarrow 0$ we obtain
 \begin{equation}
W_{\alpha =0} (x,y) =W (x,y)  = \dfrac{-1}{2\pi r^3} .
\label{eq:A-15}
\end{equation}
Finally we have showed  (\ref{eq:4-8}) for $d\geq 2$.

\section{A formula for a finite series}
In this appendix we obtain a formula for a finite series by calculating the following integral.  
\begin{equation}
A \equiv \int_{0}^\infty dk k^{2c} \int_{0}^{\pi} d\theta (\sin \theta)^{2b} e^{ikr\cos \theta -\epsilon k} 
 ~~~~ c \geq b \geq 0 ~~ b,c\in \mathbb{Z}  ~~ \epsilon ,r >0 ~~ \epsilon ,r \in \mathbb{R}  .
\label{eq:B-1}
\end{equation}
This integral is a generalization of $W_{\alpha }$ in (\ref{eq:A-1}).
The parameter $\epsilon $ and $r$ are auxiliary and they do not appear in the last formula.
We obtain the finite series when we perform the $\theta$ integral before performing the $k$ integral. 
On the other hand we obtain the simple expression when we perform the $k$ integral before performing the $\theta$ integral. 
Then we obtain the formula for the finite series. 

(i) We perform the $\theta$ integral before performing the $k$ integral.

We perform the $\theta$ integral,
\begin{equation}
 \int_{0}^{\pi} d\theta (\sin \theta)^{2b} e^{ikr\cos \theta } 
 = (1+\dfrac{1}{k^2} \dfrac{d^2}{dr^2})^b  \int_{0}^{\pi} d\theta e^{ikr \cos \theta}
 =\pi (1+\dfrac{1}{k^2} \dfrac{d^2}{dr^2})^b  J_0(kr)
\label{eq:B-2}
\end{equation}
where $J_0$ is the Bessel function of zeroth order.
We substitute (\ref{eq:B-2}) into (\ref{eq:B-1}) and perform the $k$ integral.
Then we obtain
\begin{equation}
 %=\pi \int_0^\infty dk k^a e^{-\epsilon k} (1+\dfrac{1}{k^2} \dfrac{d^2}{dr^2})^b  J_0(kr)
 A=\pi \sum_{l=0}^b {}_b \mathrm{C}_l \left( \dfrac{d}{dr} \right)^{2l} \int_0^\infty dk e^{-\epsilon k} k^{a-2l} J_0(kr)
=\pi \sum_{l=0}^b {}_b \mathrm{C}_l \left( \dfrac{d}{dr} \right)^{2l} 
\dfrac{\Gamma(\mu)}{\epsilon^{\mu}} F\left( \dfrac{\mu}{2} , \dfrac{\mu +1}{2},1;-\dfrac{r^2}{\epsilon^2} \right) 
\label{eq:B-3}
\end{equation}
where $\mu \equiv 2c-2l+1$ and $F$ is the Gaussian hypergeometric function.
We have used the condition $c\geq b\geq 0$ in the second equality in (\ref{eq:B-3}).
When $\epsilon \rightarrow 0$, we obtain
\begin{equation}
\lim_{\epsilon \rightarrow 0} \dfrac{\Gamma(\mu)}{\epsilon^{\mu}} F\left( \dfrac{\mu}{2} , \dfrac{\mu +1}{2},1;-\dfrac{r^2}{\epsilon^2} \right)
= \dfrac{\pi^{1/2}}{r^{\mu}} \dfrac{\Gamma (\mu )}{\Gamma (1-\tfrac{\mu}{2}) \Gamma (\tfrac{1+\mu}{2})}  .
\label{eq:B-4}
\end{equation}
From (\ref{eq:B-3}) and (\ref{eq:B-4}) we obtain
\begin{equation}
\lim_{\epsilon \rightarrow 0} A
= \pi^{3/2} \sum_{l=0}^b {}_b \mathrm{C}_l \dfrac{\Gamma (\mu )}{\Gamma (1-\tfrac{\mu}{2}) \Gamma (\tfrac{1+\mu}{2})} \left( \dfrac{d}{dr} \right)^{2l} \dfrac{1}{r^{\mu}}
=\dfrac{\pi^{3/2}}{r^{2c+1}} (2c)! \sum_{l=0}^b {}_b \mathrm{C}_l \dfrac{1}{\Gamma (1-\tfrac{\mu}{2}) \Gamma (\tfrac{1+\mu}{2})}  .
\label{eq:B-5}
\end{equation}

(ii) We perform the $k$ integral before performing the $\theta$ integral.

We change the variable as $t=\cos \theta$ and perform the $k$ integral,
\begin{equation}
\begin{split}
A &= \int_{-1}^{1} dt [1-t^2]^{b-1/2} \int_{0}^{\infty} dk k^{2c} e^{ikrt-\epsilon k}
= \int_{-1}^{1} dt [1-t^2]^{b-1/2} \left( \dfrac{1}{it} \dfrac{d}{dr} \right) ^{2c}  \int_{0}^{\infty} dk  e^{ikrt-\epsilon k} \\
&=i^{2c+1} \dfrac{(2c)!}{r^{2c+1} } \int_{-1}^{1} dt [1-t^2]^{b-1/2} \dfrac{1}{ (t+i z)^{2c+1}} 
\end{split}
\label{eq:B-6}
\end{equation}
where $z\equiv \epsilon /r$.
We perform the integration along the contour shown in Fig \ref{integral} (a) in the same way as Eqs.(\ref{eq:A-5})-(\ref{eq:A-7}), and obtain 
\begin{equation}
\begin{split}
\lim_{\epsilon \rightarrow 0}A &=i^{2c+1} \dfrac{(2c)!}{r^{2c+1} } (-i\pi ) \dfrac{1}{c!} (-1)^{c} (b-\tfrac{1}{2} ) (b-\tfrac{3}{2} ) \dots (b-c+ \tfrac{1}{2} )  \\
&= \pi \dfrac{(2c)!}{r^{2c+1} c!} (b-\tfrac{1}{2} ) (b-\tfrac{3}{2} ) \dots (b-c+ \tfrac{1}{2} )  \\
&= \pi \dfrac{(2c)!}{r^{2c+1} c!} \dfrac{(-1)^{c-b}}{2^c} (2b-1)!! (2c-2b-1)!! .
\label{eq:B-7}
\end{split}
\end{equation}

From $(\ref{eq:B-5}) = (\ref{eq:B-7})$ we obtain the formula for the finite series.
We simplify $(\ref{eq:B-5}) = (\ref{eq:B-7})$ and obtain the following formula,
\begin{equation}
\sum_{l=0}^{b} (-2)^{l} \dfrac{  (2c-2l-1)!! }{ l! (b-l)! (c-l)! } = (-1)^{b} \dfrac{(2b-1)!! (2c-2b-1)!!}{ b! c! }   ~~~~ c \geq b \geq 0 ~~ b,c\in \mathbb{Z}  .
\label{eq:B-8}
\end{equation}
We can also rewrite (\ref{eq:B-8}) as follows;
%The different form of the formula (\ref{eq:B-8}) is as follows;
\begin{equation}
\sum_{l=0}^{b} (-4)^{l} \dfrac{  (2c-2l)! }{ l! (b-l)! [(c-l)!]^{2} } = (-1)^{b} \dfrac{(2b)! (2c-2b)!}{ (b!)^{2} c! (c-b)! }  ~~~~ c \geq b \geq 0 ~~ b,c\in \mathbb{Z} . 
\label{eq:B-9}
\end{equation}

%\begin{equation}
%\int_0^\infty dk e^{-\epsilon k} k^{a-2l} J_0(kr)
%=\dfrac{\Gamma(\mu)}{\epsilon^{\mu}} F\left( \dfrac{\mu}{2} , \dfrac{\mu +1}{2},1;-\dfrac{r^2}{\epsilon^2} \right)
%\end{equation}

%\begin{equation}
%AB{}_{123}^{abc}C_{456}^{def}
%\end{equation}

%\cite{Bombelli:1986rw}

%\begin{acknowledgments}
%\end{acknowledgments}


%\bibliographystyle{apsrev4-1}

%\bibliography{myrefs.bib}

\begin{thebibliography}{99}
\bibitem{Bombelli:1986rw}
L. Bombelli, R. K. Koul, J. Lee, and R. D. Sorkin, Phys. Rev. D34, 373 (1986)

\bibitem{Srednicki:1993im}
 M. Srednicki, Phys. Rev. Lett. 71, 666 (1993), arXiv:hep-th/9303048

\bibitem{Hawking:2000da}
 S. Hawking, J. M. Maldacena, and A. Strominger, JHEP 05, 001 (2001), arXiv:hep-
th/0002145

\bibitem{Kabat:1995eq}
 D. N. Kabat, Nucl. Phys. B453, 281 (1995), arXiv:hep-th/9503016

\bibitem{Susskind:1994sm}
 L. Susskind and J. Uglum, Phys. Rev. D50, 2700 (1994), arXiv:hep-th/9401070

\bibitem{Frolov:1993ym}
 V. P. Frolov and I. Novikov, Phys. Rev. D48, 4545 (1993), arXiv:gr-qc/9309001

\bibitem{Jacobson:1994iw}
 T. Jacobson, (1994), arXiv:gr-qc/9404039

\bibitem{'tHooft:1984re}
 G. 't Hooft, Nucl. Phys. B256, 727 (1985)

\bibitem{Calabrese:2004eu}
 P. Calabrese and J. L. Cardy, J. Stat. Mech. 0406, P002 (2004), arXiv:hep-th/0405152

\bibitem{Holzhey:1994we}
 C. Holzhey, F. Larsen, and F. Wilczek, Nucl. Phys. B424, 443 (1994), arXiv:hep-th/9403108

\bibitem{Ryu:2006bv}
S. Ryu and T. Takayanagi, Phys. Rev. Lett. 96, 181602 (2006), arXiv:hep-th/0603001

\bibitem{Ryu:2006ef}
S. Ryu and T. Takayanagi, JHEP 08, 045 (2006), arXiv:hep-th/0605073

\bibitem{Casini:2009sr}
H. Casini and M. Huerta, J. Phys. A42, 504007 (2009), arXiv:0905.2562 [hep-th]

\bibitem{nielsen2000quantum}
 M. Nielsen and I. Chuang, Cambridge University Press, Cambridge, UK 3, 9 (2000)

\bibitem{Emig:2007cf}
T. Emig, N. Graham, R. L. Jaffe, and M. Kardar, Phys. Rev. Lett. 99, 170403 (2007),
arXiv:0707.1862 [cond-mat.stat-mech]



\bibitem{footnote}
When d=1, this is not correct because  generally the correlation function $ \bra{0} \phi (t,x) \phi(t,y) \ket{0} $ does not become zero when $|x-y|\rightarrow \infty$. 
For example the correlation function of massless free scalar fields does not become zero when $|x-y|\rightarrow \infty$.

\bibitem{footnote2}
We use the following easily verifiable identity,
\begin{equation}
\begin{pmatrix}
A & C \\
D & B 
\end{pmatrix}
= \begin{pmatrix}
A-C B^{-1} D & CB^{-1} \\
0 & 1 
\end{pmatrix}
 \begin{pmatrix}
1 & 0 \\
D & B 
\end{pmatrix}
= \begin{pmatrix}
A & 0 \\
D & 1 
\end{pmatrix}
 \begin{pmatrix}
1 & A^{-1} C \\
0 & B-DA^{-1} C 
\end{pmatrix} . \notag
\end{equation}

\end{thebibliography}

%\newpage

\end{document}